\documentclass[11pt,twoside]{article}   
\usepackage[super,sort,comma]{natbib}

\usepackage{fancyhdr}		

\usepackage{qtree}
\usepackage[labelfont={bf,it}]{caption}
\usepackage[font=small]{caption}
\newcommand{\captionv}[3]{\begin{center}\parbox{#1cm}{\caption[#2]{{\sf #3}}}
        \end{center}}

\usepackage[section]{placeins}   %

\usepackage{graphicx}

\makeatletter \renewcommand\@biblabel[1]{$^{#1}$} \makeatother
\newcommand{\cen}[1]{\begin{center} #1 \end{center}}

\usepackage[pagewise,mathlines,edtable]{lineno}

\usepackage{hyperref}
\hypersetup{ colorlinks,
    citecolor=blue,
    filecolor=blue,
    linkcolor=blue,
    urlcolor=blue
}

\usepackage{xcolor}

\definecolor{gray}{rgb}{0.6,0.6,0.6}
\definecolor{red}{rgb}{0.85,0,0}
\definecolor{green}{rgb}{0,0.85,0}
\definecolor{blue}{rgb}{0,0,0.85}
\definecolor{beige}{rgb}{0.92,0.87,0.78}
\usepackage[all]{hypcap}    

\usepackage{subcaption} 
\usepackage{algorithm}
\usepackage{program}
\usepackage{multirow}

\usepackage{setspace}
\usepackage{amsmath}
\DeclareMathOperator*{\argmax}{argmax}

\usepackage{algpseudocode}

\makeatletter
\def\therule{\makebox[\algorithmicindent][l]{\hspace*{.5em}\vrule height 0.75\baselineskip depth .25\baselineskip}}%

\newtoks\therules
\therules={}
\def\appendto#1#2{\expandafter#1\expandafter{\the#1#2}}
\def\gobblefirst#1{
  #1\expandafter\expandafter\expandafter{\expandafter\@gobble\the#1}}%
\def\LState{\State\unskip\the\therules}
\def\pushindent{\appendto\therules\therule}%
\def\popindent{\gobblefirst\therules}%
\def\printindent{\unskip\the\therules}%
\def\printandpush{\printindent\pushindent}%
\def\popandprint{\popindent\printindent}%

\algdef{SE}[WHILE]{While}{EndWhile}[1]
  {\printandpush\algorithmicwhile\ #1\ \algorithmicdo}
  {\popandprint\algorithmicend\ \algorithmicwhile}%
\algdef{SE}[FOR]{For}{EndFor}[1]
  {\printandpush\algorithmicfor\ #1\ \algorithmicdo}
  {\popandprint\algorithmicend\ \algorithmicfor}%
\algdef{S}[FOR]{ForAll}[1]
  {\printindent\algorithmicforall\ #1\ \algorithmicdo}%
\algdef{SE}[LOOP]{Loop}{EndLoop}
  {\printandpush\algorithmicloop}
  {\popandprint\algorithmicend\ \algorithmicloop}%
\algdef{SE}[REPEAT]{Repeat}{Until}
  {\printandpush\algorithmicrepeat}[1]
  {\popandprint\algorithmicuntil\ #1}%
\algdef{SE}[IF]{If}{EndIf}[1]
  {\printandpush\algorithmicif\ #1\ \algorithmicthen}
  {\popandprint\algorithmicend\ \algorithmicif}%
\algdef{C}[IF]{IF}{ElsIf}[1]
  {\popandprint\pushindent\algorithmicelse\ \algorithmicif\ #1\ \algorithmicthen}%
\algdef{Ce}[ELSE]{IF}{Else}{EndIf}
  {\popandprint\pushindent\algorithmicelse}%
\algdef{SE}[PROCEDURE]{Procedure}{EndProcedure}[2]
   {\printandpush\algorithmicprocedure\ \textproc{#1}\ifthenelse{\equal{#2}{}}{}{(#2)}}%
   {\popandprint\algorithmicend\ \algorithmicprocedure}%
\algdef{SE}[FUNCTION]{Function}{EndFunction}[2]
   {\printandpush\algorithmicfunction\ \textproc{#1}\ifthenelse{\equal{#2}{}}{}{(#2)}}%
   {\popandprint\algorithmicend\ \algorithmicfunction}%
\makeatother

\graphicspath{{}{../Figures/}} 

\usepackage{subfiles} 

\usepackage{fancyhdr,graphicx,amssymb}
\algrenewcommand\alglinenumber[1]{\tiny #1:}
\begin{document}

\cen{\sf {\Large {\bfseries A reinforcement learning application of guided Monte Carlo Tree Search algorithm for beam orientation selection in radiation therapy } \\  
\vspace*{10mm}
Azar Sadeghnejad-Barkousaraie, Gyanendra Bohara, Steve Jiang, Dan Nguyen*} \\
\vspace*{10mm}
Medical Artificial Intelligence and Automation (MAIA) Laboratory, Department of Radiation Oncology, UT Southwestern Medical Center, Dallas, TX
\vspace{5mm}\\
}

\pagenumbering{roman}
\setcounter{page}{1}
\pagestyle{plain}
* corresponding author: Dan.Nguyen@UTSouthwestern.edu \\

\begin{abstract}
\noindent {\bf Purpose:} Due to the large combinatorial problem, current beam orientation optimization algorithms for radiotherapy, such as column generation (CG), are typically heuristic or greedy in nature, leading to suboptimal solutions. We propose a reinforcement learning strategy using Monte Carlo Tree Search capable of finding a superior beam orientation set and in less time than CG.\\
{\bf Methods:} We utilized a reinforcement learning structure involving a supervised learning network to guide Monte Carlo tree search (GTS) to explore the decision space of beam orientation selection problem. We have previously trained a deep neural network (DNN) that takes in the patient anatomy, organ weights, and current beams, and then approximates beam fitness values, indicating the next best beam to add. This DNN is used to probabilistically guide the traversal of the branches of the Monte Carlo decision tree to add a new beam to the plan. To test the feasibility of the algorithm, we solved for 5-beam plans, using 13 test prostate cancer patients, different from the 57 training and validation patients originally trained the DNN. To show the strength of GTS to other search methods, performances of three other search methods including a guided search, uniform tree search and random search algorithms are also provided. \\
{\bf Results:} On average GTS outperforms all other methods, it find a solution better than CG in 237 seconds on average, compared to CG which takes 360 seconds, and outperforms all other methods in finding a solution with lower objective function value in less than 1000 seconds. Using our guided tree search (GTS) method we were able to maintain a similar planning target volume (PTV) coverage within 1$\%$ error, and reduce the organ at risk (OAR) mean dose for body, rectum, left and right femoral heads, but a slight increase of $1.\%$ in bladder mean dose. \\
{\bf Conclusions:} In this study we demonstrate that our GTS method produces superior plans compared to CG method, and in shorter time, and therefore is more suitable for clinical application.
\end{abstract}
\newpage     



\setlength{\baselineskip}{0.7cm}      

\pagenumbering{arabic}
\setcounter{page}{1}
\pagestyle{fancy}
\section{Introduction}
Radiation therapy is one of the main modalities to cure cancer, and is used in over half of cancer treatments, either standalone or in conjunction with another modality, such as surgery or chemotherapy. For intensity-modulated radiation therapy (IMRT), the patient body is irradiated from fixed beam locations around patient body, and the radiation field is modulated at each beam position using multi-leaf collimators (MLC). In IMRT, the optimal choice of beam orientations has a direct impact on the treatment plan quality, influencing the final treatment outcome, hence patient quality of life. Current clinical protocols either have the beam orientations selected by protocol or manually by the treatment planner. Beam orientation optimization (BOO) methods solve for a suitable set of beam angles by solving an objective function to a local minimum. BOO has been studied extensively in radiation therapy procedures, for both coplanar \citep{Bortfeld1993OptimizationConsiderations,Yan1999,Pugachev2001,Djajaputra2003,Li2004AutomaticAlgorithm,Li2005,Romeijn2005A,Schreibmann2005,Aleman2008,Lim2008,Breedveld2009,Lim2009APlanning,Craft2010,Bangert2010,Breedveld2012,Rocha2013BeamMethod,Yuan2015a,Amit2015,Liu2017a,Cabrera-Guerrero2018ComparingRadiotherapy,Rocha2018,OConnor2018,Cabrera-Guerrero2018,CabreraG.2018}, and noncoplanar  \citep{Pugachev2001,Djajaputra2003,Potrebko2008,Llacer2009,Bangert2010,Breedveld2012,Liu2017a,Yu2018,Yarmand2018EffectiveTherapy,OConnor2018,Yuan2018,Rocha,Bedford2019,Ventura2019ComparisonIMRT} IMRT, or intensity-modulated proton therapy\citep{Oelfke2001,Gu2018,Shirato2018SelectionTreatment,Gu2019} (IMPT) by researchers in the past three decades. However, BOO has not been widely adopted due to their high computational cost and complexity, since it is a large-scale NP hard combinatorial problem\citep{Azizi-Sultan2006,Yuan2018}. Despite the extensive research, the lack of practical clinically beam orientation selection algorithms still exists due to the computational and time intensive procedure, as well as the sub-optimality of the final solution, and BOO remains a challenging step of the treatment planning process. 

To measure the quality of the BOO solution, it is necessary to calculate dose influence matrices of each potential beam orientation. Dose influence matrices for one beam associates all the individual beamlets in the fluence map with the voxels of the patient body. This calculation is time consuming and requires a large amount of memory to use in optimization. To mange the limited capacity of computational resources, the treatment planning process, after defining the objective function, is divided into two major steps: 1) find a suitable set of beam orientations, and 2) solve the fluence map optimization problem (FMO)\citep{CabreraG.2018} of those selected beams. However, these two steps are not independent of each other--the quality of BOO solution can be evaluated only after FMO is solved, and FMO can be defined only after BOO is solved.  Due to the non-convexity and large scale of the problem, researchers consider dynamic programming methods by breaking the problem into a sequence of smaller problems. One of the successful algorithms specially for solving complex problems such as BOO is a method known as Column Generation (CG). In the original application of CG into radiotherapy, \citet{Romeijn2005AModulation} solved a direct aperture optimization (DAO) problem by using CG. \citet{Dong2013} then proposed a greedy algorithm based on column generation, which iteratively adds beam orientations until the desired number of beams are reached. \citet{Rwigema20154Toxicity} use CG to find a set of 30 beam orientations to be used in $4\pi$ treatment planning of stereotactic body radiation therapy (SBRT) for patients with recurrent, locally advanced, or metastatic head-and-neck cancers, to show the superiority of $4\pi$ treatment plans to those created by volumetric modulated arc therapy (VMAT). \citet{Nguyen2016} used CG to solve the triplet beam orientation selection problem specific to MRI guided Co-60 radiotherapy. \citet{Yu2018} used an in-house CG algorithm to solve an integrated problem of beam orientation and fluence map optimization.  

However, CG is a greedy algorithm that has no optimality guarantee, and typically yields a sub-optimal problem. In addition, CG still takes as much as 10 minutes to suggest a 5 beam plan for prostate IMRT. The aim of this work is to find a method to explore a larger area of the decision space of BOO in order to find higher quality solutions than that of CG in a short amount of time. The proposed method starts with a deep neural network that has been trained using CG as a supervisor. This network can mimic the behavior of CG by directly learning CG's fitness evaluations of the beam orientations in a supervised learning manner. The efficiency of this supervised network, which can propose a set of beam angles that are non-inferior to that of CG, within less than two seconds, is presented in our previous work\citep{SadeghnejadBarkousaraie2019ATherapy}. Given a set of already selected beams, this network will predict the fitness value of each beam, which is how much the beam will improve the objective function when added in the next iteration.

In this study, we extend our previous work, and combine this trained supervised learning (SL) network with a reinforcement learning method, called Monte Carlo tree search. We use these fitness values from the SL network as a guidance to efficiently navigate action space of the reinforcement learning tree. Specifically, it provides the probability of selecting a beam in the search space of the tree at each iteration, so that the beam with the better likelihood to improve the objective function has the higher chance of being selected at each step. To evaluate our proposed method, we compare its performance against the state-of-the-art CG. We developed three additional combinations of the guided and random search tree approaches for comparison. 
\section{Methods}
The proposed method has a reinforcement learning structure involving a supervised learning network to guide Monte Carlo tree search to explore the beam orientation selection decision space. This method, guided Monte Carlo tree search (GTS), consists of two main phases: 1) Supervised training a deep neural network (DNN) to predict the probability distribution of adding the next beam, based on patient anatomy, and 2), using this network for a guided Monte Carlo tree search method to explore a larger decision space more efficiently to find better solutions. For the first phase we use the CG implementation for BOO problem, where CG iteratively solves a sequence of Fluence Map Optimization (FMO) problems \citep{CabreraG.2018} by using GPU-based Chambolle-Pock algorithm \cite{Chambolle2010}, the results of the CG method are used to trained a supervised neural network. For the second phase, which is the main focus of this work, we present a Monte Carlo Tree Search algorithm, using the trained DNN. Each of these phases are presented in the following sections. 

\FloatBarrier\subsection{Supervised Learning of the Deep Neural Network\label{DNN-training}}

\begin{figure}
\centering
    \includegraphics[width=1.\textwidth]{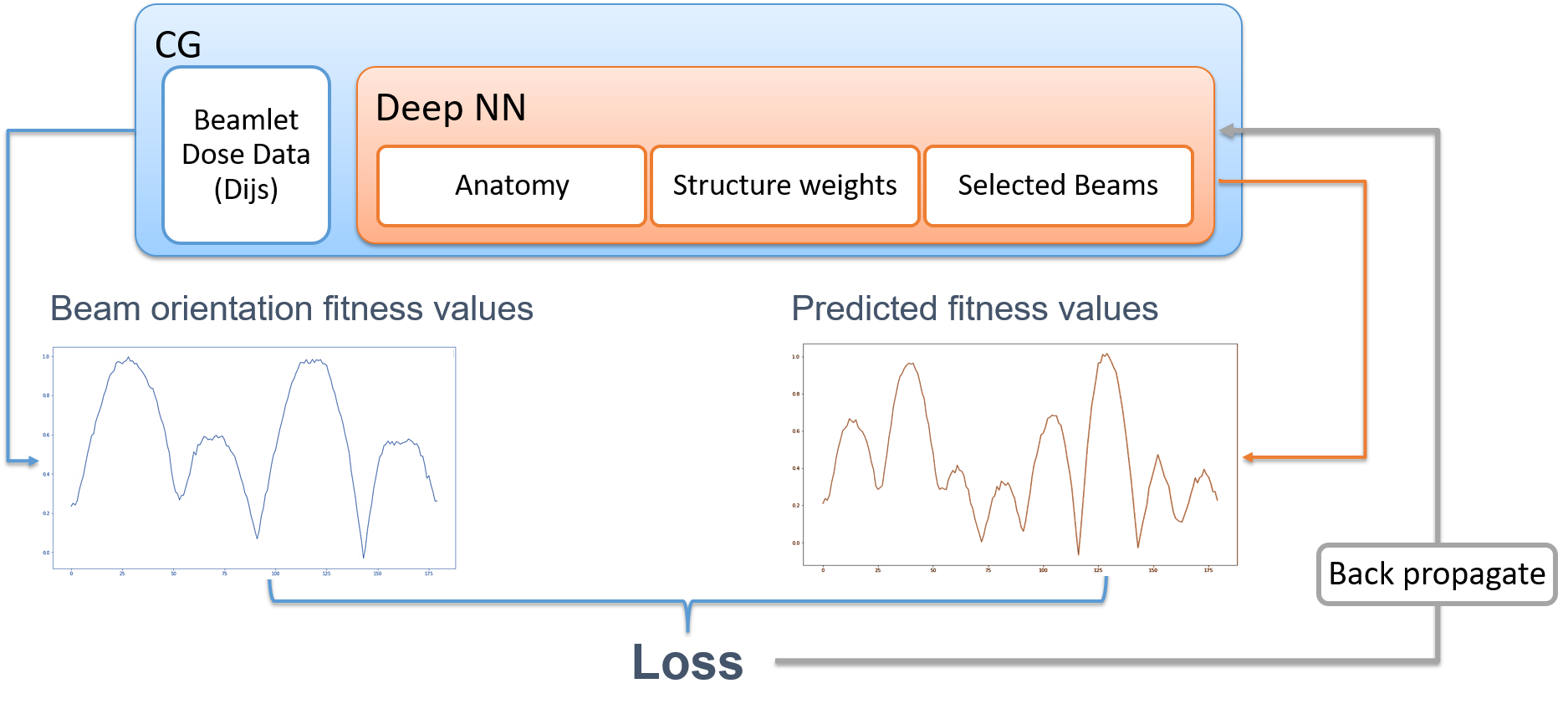}
    \captionv{15}{Supervised-structure}{Schematic of the Supervised Training Structure to predict Beam Orientation fitness values. Column Generation (CG) as teacher and deep neural network (DNN) as Trainee.
    \label{fig:supervisedStructure}}
\end{figure}

We develop a deep neural network (DNN) model that learns from column generation how to find fitness values for each beam based on the anatomical features of a patient and a set of structure weights for the planning target volume (PTV) and organs-at-risk (OAR). The CG greedy algorithm starts with an empty set of selected beams, calculates the fitness values of each beam based on the optimality condition of the objective function shown in \autoref{main-obj}.
\begin{equation}
\min_{x} \frac{1}{2}\sum_{\forall s \in S}w_s^2 \|D_{s}x - p\|_2^2 \quad  s.t.\: x \geq 0    \label{main-obj}
\end{equation}
where $w_s$ is the weight for structure s, which are pseudo randomly generated between zero and one during the training process to generate many different scenarios. The value, $p$, is the prescription dose for each structure, which is assigned 1 for the PTV and 0 for OARs. At each iteration of CG, fitness values are calculated based on Karush–Kuhn–Tucker (KKT) conditions\citep{kuhn1951nonlinear,Karush2014} of a master problem, and they represent how much improvement each beam can make in the objective function value. The beam with the highest fitness value is selected to be added to the selected beam set, $S$. Then, FMO for the selected beams is performed, which affects the fitness value calculations for the next iteration. The process is repeated until the desired number of beams are selected. The supervised DNN learns to mimic this behavior through the training of the DNN is shown in figure \ref{fig:supervisedStructure}. Once trained, this DNN is capable of efficiently providing a suitable set of beam angles in  less than 2 seconds, as opposed to the 360 seconds required to solve the same problem using CG. The details of the DNN structure and its training process is described in our previous work \cite{SadeghnejadBarkousaraie2019ATherapy}.

Patient anatomical features include the contoured structures (organs at risk) of the images from patients with prostate cancer and the treatment planning weights assigned to each structure. The images of 70 prostate cancer patients are used for this research, each with 6 contours: planning target volume (PTV), body, bladder, rectum, and left and right femoral heads. From 70 patients, 50 was randomly selected to train the network and 7 for its validation. The remaining 13 patients images is used for testing and applying the Monte Carlo tree search method. 

 \FloatBarrier\subsection{Monte Carlo Tree Search}
The pre-trained DNN probabilistically guides the traversal of the branches on the Monte Carlo decision tree to add a new beam to the plan. Each branch of the tree starts from root as an empty set of beams, and continues until it reaches the terminal state. After the exploration of each complete plan (selection of 5 beams in our case), the fluence map optimization problem is solved and and based on that, the probability distribution to select next beam will be updated, using the reward function, in the backpropagation stage. Then, starting from root the exploration of the next plan will begin until the stopping criteria is met. \autoref{fig:tree_structure} shows an example of a tree search, which has discovered seven plans so far. 

\begin{figure}
    \qtreecentertrue
    \Tree[.Root [.${b_1}^{1}$ [.${b_2}^{2}$ [.${b_3}^{3}$ [.${b_4}^{4}$ ${b_5}^{5}$ ]]]
                      [.${b_6}^2$ [.${b_7}^3$ [.${b_8}^4$ ${b_9}^5$ ]
                               [.${b_{10}}^5$ ${b_{11}}^5$ ] 
                               ]]]
          [.${b_{12}}^1$ 
                [.${b_{13}}^2$ [.${b_{14}}^3$ [.${b_{15}}^4$ ${b_{16}}^5$ ]
                                              [.${b_{20}}^4$ ${b_{21}}^5$ ${b_{22}}^5$ ]]
                               [.${b_{17}}^3$ [.${b_{18}}^4$ ${b_{19}}^5$ ]]
                                                      ]]]
    \captionv{15}{Short title - can be blank}
    {An example of guided tree search, subscript are order that a node is generated, and superscript is the depth of the node in the tree.
     \label{fig:tree_structure}}
\end{figure}
\subsubsection{Basics of Monte Carlo Tree Search}
 Monte Carlo Tree Search (MCTS) uses the decision tree to explore the decision space, by randomly sampling from it\cite{Browne2012}. The search process of MCTS consists of four steps: 1) node selection, 2) expansion, 3) simulation, and 4) back-propagation on the simulation result. To explain these processes in detail, there are some properties that need to be defined first, these definitions are as follows:
 \begin{description}
     \item[State of the problem:] include patient’s anatomical features and a set of selected beam orientations ($B$). At the beginning of the planning, this set has no member, and it is updated throughout the solution procedure.
    \item[Actions:] the selection of the next beam orientation to be added to set B, given the state of the problem.
    \item[Solution or terminal state:] state of the problem in which the number of selected beam orientations (size of $B$) is the same as a predefined number of beams ($N$), chosen by user. At this point, a feasible solution for the problem is generated. 
     \item[Decision Tree:] The solution space of a set of discrete numbers--in this work discrete numbers are the beam orientations--specially with iterative structures, can be defined as a tree, where each node and branch represent the selection of one number or a subset of available numbers, respectively.
    \item[Node ($Y$):] selection of one potential beam orientations is a node.
    \item[Root ($O$):] a node with empty set of beam orientations, every solution start from the root.
    \item[Path:] a unique connected sequence of nodes in a decision tree.
    \item[Branch ($Q$):] a path originated from Root node. Each branch represents the iterative structure of the proposed method. The length of a branch is the number of nodes in that branch. In this work solution is a branch with size $N+1$. There is only one branch from each root to any node in a tree.
    \item[Leaf:] last node of a branch. There is no exploration of the tree after a leaf is discovered.
    \item[Internal node:] any node in a tree except for root and leaves.
\end{description}
The \textbf{selection} process in the proposed method is guided by a pre-trained DNN as described in \autoref{DNN-training}. This DNN is used to probabilistically guide the traversal of the branches on the Monte Carlo decision tree to add a new beam to the plan. At each node--starting by root note--the DNN is called to predict an array of fitness values for each beam orientation($P$). An element of this array $P[i]$ represents the likelihood of the selection of the $i^{th}$ beam orientation. For example, if the number of potential beam orientations is 180, with $2^{\circ}$ separation, $Y$ would be an array of size 180, and $P[2]$ is the likelihood of selecting beam orientation in $2^{nd}$ position of the potential beam orientations, $P[2]=4^\circ$.
The \textbf{expansion}  process happens after selection process at internal nodes, to further explore the tree and create children nodes. The traversal approach in the proposed method is depth first, which means that the branch of a node, that is visited or created for the first time, continues expanding until there are $N+1$ nodes in a branch. In this case, selection and expansion processes are overlapping because only one child node is created or visited at a time, although a node can be visited multiple times and several children can be generated from one node, except for leaf. The leaf node does not have any children. Nodes in a branch must be unique, it means that a branch of each external node ($Q$) can be expanded only to nodes that are not already in the branch. 
In fact, beam orientation optimization problem can be defined as a dynamic programming problem with the following formula:
\begin{equation}
    G_{k}^{S}=S\union\{k\}\union{G_{n^*}^{S\union\{k\}}} \mid n^*=\argmax_{n>k}V_{G_n^{S\union\{k\}}}\label{dynamic}
\end{equation}
where $S$ is a set of indices for previously selected beams, $k$ is index of currently selected beam and ${G_{n^*}^{S\union\{k\}}}$ is a subset of beams to be selected that has the highest reward value. 
 Each $\bf{simulation}$ consists of iteratively selecting a predefined number of beams ($N$), in this work $N=5$. After the exploration of each complete plan, the fluence map optimization problem is solved and used for the \textbf{back-propagation} step, which is used to update the probability distribution for beam selection. 
 
 \subsubsection{Main Algorithm}
 The detailed of the guided Monte Carlo tree search algorithm in the form of a pseudo code is provided in Algorithm \ref{alg:GTS}. Several properties of each node in the proposed tree are being updated after the exploration of final states. To simplify the algorithm, these properties are addressed as variables and the following is a list of them:
\begin{description}
    \item[Cost ($Y_v$):] After a leaf is discovered, an FMO problem associated with the beams of that branch will be solved, the value of the FMO cost function is the value associated with its corresponding leaf. The cost value of all other nodes (other than leaves) in a tree is the average cost of its sub-branches. For example in \autoref{fig:tree_structure} the cost value of node ${b_1}^1$ is the average cost of nodes ${b_2}^2$ and ${b_6}^2$.
    \item[probability distribution ($Y_P$):] an array of size 180 (the number of potential beam orientations), where $i^{th}$ element of this array represents the chance of improvement in the current cost value if tree branches out by selecting $i^{th}$ beams. After a node is discovered in the tree, this distribution is populated by using DNN. After the first discovery of a node, $Y_P$ is updated based on the reward values.
    \item[Reward ($Y_R$):] is a function of the node's cost values and the best cost value ever discovered in the search process. The reward values would be updated after each cost calculation and are calculated and updated by the reward calculation procedure defined in line \ref{Reward_function} of Algorithm \ref{alg:GTS}.
    \item[Depth ($Y_D$):] is simply the number of beam orientations selection after node $Y$ is discovered.
    \item[Name ($Y_{id}$):] a unique string value as id for each node, this value is the path from root to node $Y$.
    \item[Beam Set ($Y_B$):] the set of beams selected for a branch started from root and ended in node $Y$.
    \item[Parent ($Y_{parent}$):] the immediate node before node $Y$ in a branch from root to $Y$, except for root node, all other nodes in a tree have one parent.
    \item[Children ($Y_{children}$)]: the immediate node(s) of the sub-branches from the node $Y$, except for leaves, all other nodes in a tree have at-least one children.
\end{description}

\begin{algorithm}[htbp]
    \footnotesize
    \caption{\textsc{Select $N$ beam orientations from $M$ candidate beams}}
    \label{alg:GTS}
    \begin{algorithmic}[1]
    \Procedure{Initialization}{}
        \LState{set selected beam as $B\gets\emptyset$ an empty set, best cost value as infinity (${V^*}\gets{\infty}$), and best selected beam as ${B^*}\gets{\emptyset}$} 
        \LState{create a root node object ($O$) with the following properties:}
        \LState{\ \ $\qquad$  name($O_{id}\gets{\mathtt{Root}}$), probability distribution($O_P\gets{\emptyset}$), number of visits($O_Z\gets{0}$), beam index($O_b$)} 
        \LState{\ \ $\qquad$  reward($O_R\gets{0}$), cost($O_V\gets{\infty}$), depth($O_D\gets{0}$), parent($O_{parent}\gets{\emptyset}$), children($O_{children}\gets{\emptyset}$)}
        \LState {assign root node to current node(${Y^{\#}}\gets{O}$)} 
        \LState {given the set $B$ as input to $\mathtt{DNN}$, predict an array of fitness values and assign it to root node ${Y^{\#}}_P\gets{Prd(\mathtt{DNN},B)}$} 
        \LState {set  $stop \gets{False}$}
    \EndProcedure
    \While{$stop$ is ${False}$}
    	\LState {choose the next beam index ($b$) using the probability distribution of the current node ${Y^{\#}}_{P}$}  
    	\LState  {create string name as ${ID} \gets{{Y^{\#}}_{id} \concat {b}}$} \Comment{$\concat$: string concatenation}
    	\LState {update selected beam set $B\gets{B\cup{b}}$ }
    	\If  {$ID \notin {{Y^{\#}}_{children}}_{id}$ (current node does not have a child named ${ID}$)}
    	    \LState {create a new node $Y$ with ${Y}_{id}\gets{ID}$}
    	    \LState {${{Y}_{parent}}\gets{Y^{\#}}$, ${{Y}_{Z}}\gets 1$}
    	    \LState {predict an array of fitness values $F(\mathtt{DNN},B)$} 
    	    \LState {assign predicted values to new node ($Y_P\gets{Prd(\mathtt{DNN},B)}$)}
    	    \LState {$Y_D\gets{{Y^{\#}}_D + 1}$}
    	    \LState {set beam index($Y_b\gets{b}$)}
    	    \LState {add $Y$ as a new child ${Y^{\#}}_{children}\gets{{Y^{\#}}_{children} \cup Y}$ }
    	    \LState {update current node $Y^{\#}\gets{Y}$}
    	\Else
    	    \LState {update current node ($Y^{\#}\gets{X \vert\{ X \in {Y^{\#}}_{children} \& X_{id}=ID}\}$)}
    	    \LState {update visit parameter of current node, (${Y^{\#}}_Z\gets{{Y^{\#}}_Z + 1}$)}
        \EndIf
        \If{$\vert B\vert = N$ or ${Y^{\#}}_D = N$ }
            \Procedure{Reward Calculations \label{Reward_function}}{}
            \LState {solve $FMO$ given set $B$ and save as ${Y^{\#}}_V\gets{Fmo(B)}$}
            \If {${V^*} > {Y^{\#}}_V$}
                \LState {set ${V^*} \gets {Y^{\#}}_V$, {${B^*}\gets B$}, {${Y^{\#}}_R\gets 1.$}}
            \Else:
                \LState {${Y^{\#}}_R \gets ({{V^*}-{Y^{\#}}_V})/{{Y^{\#}}_V} +0.15$ }
            \EndIf
            \While{${Y^{\#}}_{id}\ne \mathtt{Root}$}
                \LState{$Y^{\#}\gets{{Y^{\#}}_{parent}}$}
                \LState{${Y^{\#}}_R \gets{{\sum_{{X}\in {Y^{\#}}_{children}}{{X_R}}}/{\vert {Y^{\#}}_{children}\vert}}$}
                \For {$X\in {Y^{\#}}_{children}$}
        	        \LState {${Y^{\#}}_D[X_b]\gets{{Y^{\#}}_D[X_b]/X_Z +\sqrt{\ln{X_Z}/{Y^{\#}}_Z}}$}
                    \If {${Y^{\#}}_V > X_V$}
                        \LState {${Y^{\#}}_V\gets{X_V}$}
        	        \EndIf
                \EndFor
            \EndWhile
            \EndProcedure
            \If {stopping criteria is met}
                \LState {$Stop \gets{True}$}
            \Else:
                \LState {$B \gets{\emptyset} , Y^{\#}\gets{O}$}
            \EndIf
        \EndIf
    \EndWhile
    \LState {\Return ${{V^*},{B^*}}$}
    \end{algorithmic}
    \vspace{8pt}
\end{algorithm}

\subsection{Algorithms for performance comparison \label{fouralgorithms}}
In general, four frameworks were designed to show the efficiency of the proposed GTS method compared to others. These models are defined as:
\begin{description}
\item[Guided Tree Search (GTS)]: As presented in Algorithm \ref{alg:GTS}, used a pre-trained policy network to guild a Monte-Carlo decision tree.
\item[Guided Search (GuidS)]: Used the pre-trained network to search the decision space by iteratively choosing one beam based on the predicted probabilities from the policy network. Unlike GTS, the search policy is not updated as the search progresses. This process is detailed in Algorithm \ref{alg:GuidS}. 
\item[Randomly sample Tree Search (RTS)]: This algorithm is simple Monte-Carlo tree search method which starts with a uniform distribution to select beam orientations (randomly select them), and then update the search policy as the tree search progresses. Note that all of tree operations used in GTS is also used in this algorithm, except for having a policy network to guide the tree. This method is proposed to show the impact of using DNN to guild the decision tree.
\item[Random Search (RandS)]: This method searches the decision space with uniformly random probability until stopping criteria is met. It randomly selects 5 beam orientations and solves the corresponding FMO problem. The search policy is not updated. Its procedure is close to Algorithm \ref{alg:GuidS} where the ``Select $B$ using DNN'' procedure is replaced by randomly selecting 5 unique beams.
\end{description}

\begin{algorithm}[tbp]
\footnotesize
\caption{\textsc{Guided Search algorithm(GuidS)}}
\label{alg:GuidS}
\hspace*{\algorithmicindent} \textbf{Input:}{ Pre-trained DNN}
    \begin{algorithmic}[1]
        \LState {initialize $B$ as an empty array, best cost value as infinity (${V^*}\gets{\infty}$), and best selected beam as ${B^*}\gets{\emptyset}$}
        \LState {set current number of beam orientations in $B$ as 0, $N_B \gets{0}$}
        \LState {set $stop \gets{False}$}
        \While{$stop$ = $False$}
            \Procedure{Select $B$ using DNN}{}
            \While{$N_B < N$}
            	\LState{predict an array of fitness values $P=F(\mathtt{DNN},B)$}
            	\LState{Select next node ($b$) with the probability of $P(b)$}
            	\LState{Update $B$: $B=B\union\{b\}$ and $N_B = N_B +1$}
            \EndWhile
            \LState \Return{$B$}
            \EndProcedure
            \LState {solve $FMO$ given set $B$ and save as ${V\gets{Fmo(B)}}$}
            \If{$V<{V^*}$}
                \LState{${V^*} \gets{V}$ and ${B^*}\gets{B}$}
            \EndIf
            \If {stopping criteria is met}
                \LState {$stop \gets{True}$}
            \Else:
                \LState {$B \gets{\emptyset} , N_B \gets{0}$}
            \EndIf
        \EndWhile
        \LState{\Return{${B^*}$ and ${V^*}$}}
    \end{algorithmic}
    \vspace{2pt}
\end{algorithm}

\FloatBarrier\subsection{Data}
We used images from 70 patients with prostate cancer, each with 6 contours: PTV, body, bladder, rectum, left femoral head, and right femoral head. Additionally, the skin and ring tuning structures were added during the fluence map optimization process to control high dose spillage and conformity in the body. The patients were divided randomly into two exclusive sets: 1) a model development set, which includes training and validation data, consisting of 57 patients, 50 for training and 7 for validation, for cross-validation method, and 2) a test data set consisting of 13 patients. 
Each patient’s data contains 6 contours: PTV, body, bladder, rectum, and left and right femoral heads. Column generation was implemented with a GPU-based Chambolle-Pock algorithm\citep{Chambolle2010}, a first-order primal-dual proximal-class algorithm, to create 6270 training and validation scenarios (22 5-beam plans for each of 57 patients) and 130 test scenarios (10 5-beam plans for each of 13 test patients). The DNN trained over 400 epochs, each with 2500 steps and batch size of one. 

The performances of four methods GTS, GuidS, RTS, and RandS, explained in section \autoref{fouralgorithms}, are evaluated. Two of these methods, GTS and GuidS, use the pre-trained DNN as a guidance network. We originally had the images of 70 patients with prostate cancer, and used the images of the 57 of them to train and validate DNN and therefore cannot be used for the testing in this project\footnote{ To keep the proposed methods completely independent from the dataset used for training DNN.}. There are 13 patients that DNN has never seen before and the images of those patients are used in this project as test set. Multiple scenarios can be generated for each patient, based on the weights assigned to patient's structures for planning their treatments. We semi-randomly generated 10 sets of weights for each patients. In total, we have a total of 130 test plans among the 13 test patients for the comparison. All the tests in this paper were performed on a computer with an Intel Core I7 processor@3.6 GHz, 64 GB memory, and an NVIDIA GeForce GTX 1080 Ti GPU with 11 GB video memory.

The structure weight selection scheme is outlined by the following process:
\begin{enumerate}
    \item{ In $50\%$ of the times, a uniform distribution in the range of 0 to 0.1 is used to generate a weight for each OAR separately.} 
    \item {	In $10\%$ of the times, the smaller range of 0 to 0.05 is used to select weights for OARs separately, with uniform distribution.}
    \item {	And finally in $40\%$ specific ranges were used for each OAR: Bladder: [0,0.2], Rectum: [0,0.2], Right Femoral Head: [0,0.1], Left Femoral Head: [0,0.1], Shell: [0,0.1] and Skin: [0,0.3]}
\end{enumerate}
The weights range from 0 to 1. This weighting scheme was found to give a clinically reasonable dose, however, the dose itself may not be approved by the physician for that patient. 

Finally, considering only test scenarios, FMO solutions of beam sets generated by CG and by the 4 tree search methods were compared with the following metrics: 
\begin{description}
    \item [PTV $\bf{{D}_{98}}$, PTV $\bf{{D}_{99}}$:]{The dose that $98\%$ and $99\%$, respectively, of the PTV received}
	\item [PTV $\bf{{D}_{max}}$:]{Maximum dose received by PTV, the value of $D_2$ is considered for this metric}
	\item [PTV Homogeneity:]{$\frac{{{PTV} {D_2}} - {PTV D_{98}}}{{PTV} {D_{50}}}$ where $PTV D_2$ and $D_{50}$ are the dose received by $2\%$ and $50\%$, respectively, of PTV}
	\item [ Paddick Conformity Index (${CI}_{Paddick}$) \citep{van1997conformation,paddick2000simple}:] {$\frac{{(V_{PTV}  \cap V_{100\%Iso})}^2}{V_{PTV} \times V_{100\%Iso}}$ where $V_{PTV}$ is the volumne of the PTV and $V_{100\%Iso}$  is the volume of the isodose region that received $100\%$ of the dose}
	\item [High Dose Spillage ($\bf{R_{50}}$):] {$\frac{V_{50\%Iso}}{V_{PTV}}$  where $V_{50\%Iso}$ is the volume of the isodose region that received $50\%$ of the dose}
\end{description}

\begin{figure}[tb]
    \begin{subfigure}{.47\textwidth}
        \centering
        \includegraphics[clip,trim= 2mm 4mm 85mm 3mm,scale=0.178]{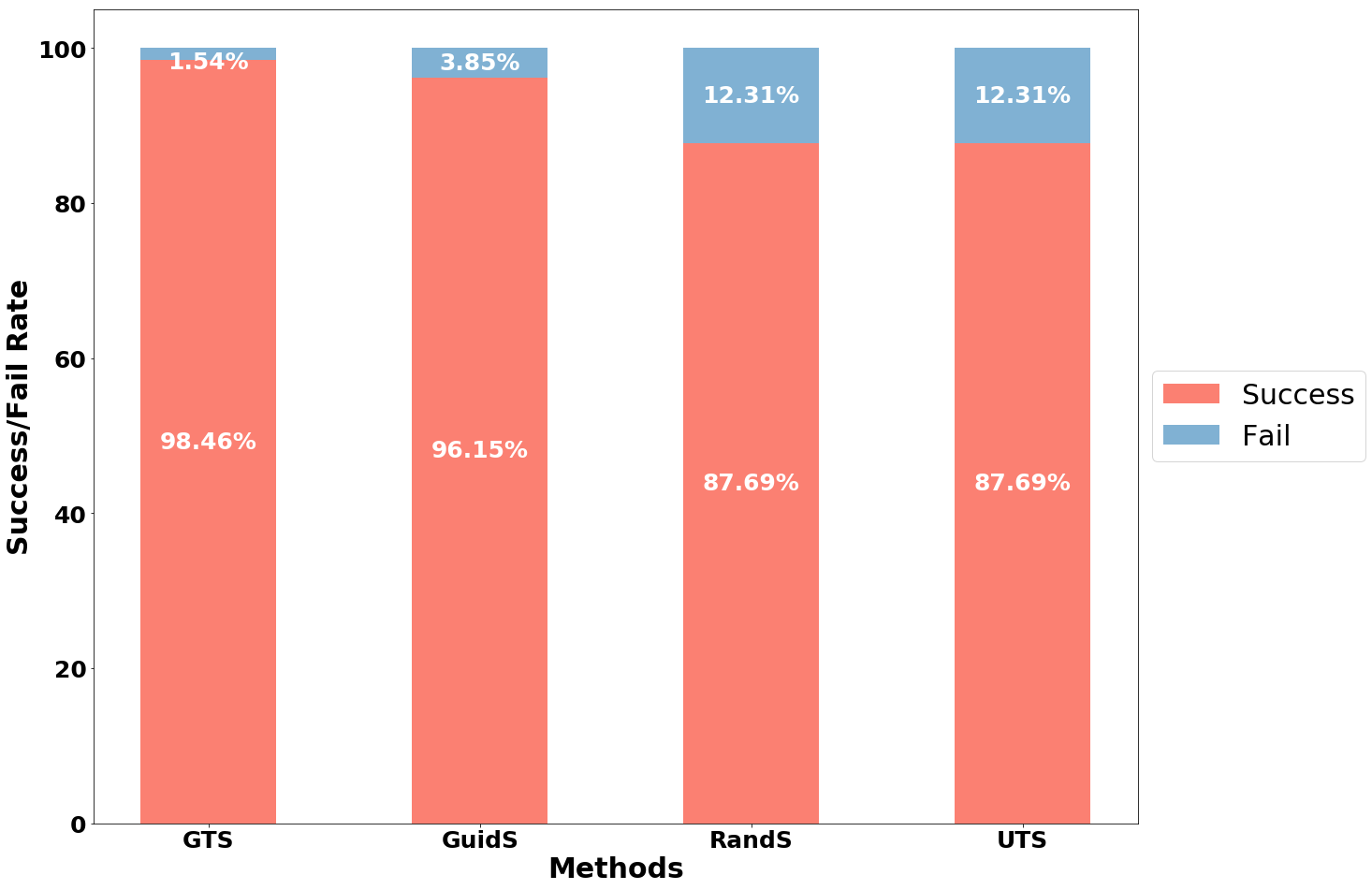}
        \caption{
        \label{fig:success-fail}}
    \end{subfigure}
    \begin{subfigure}{.52\textwidth}
        \centering
        \includegraphics[clip,trim= 35mm 4mm 2mm 3mm,scale=.178]{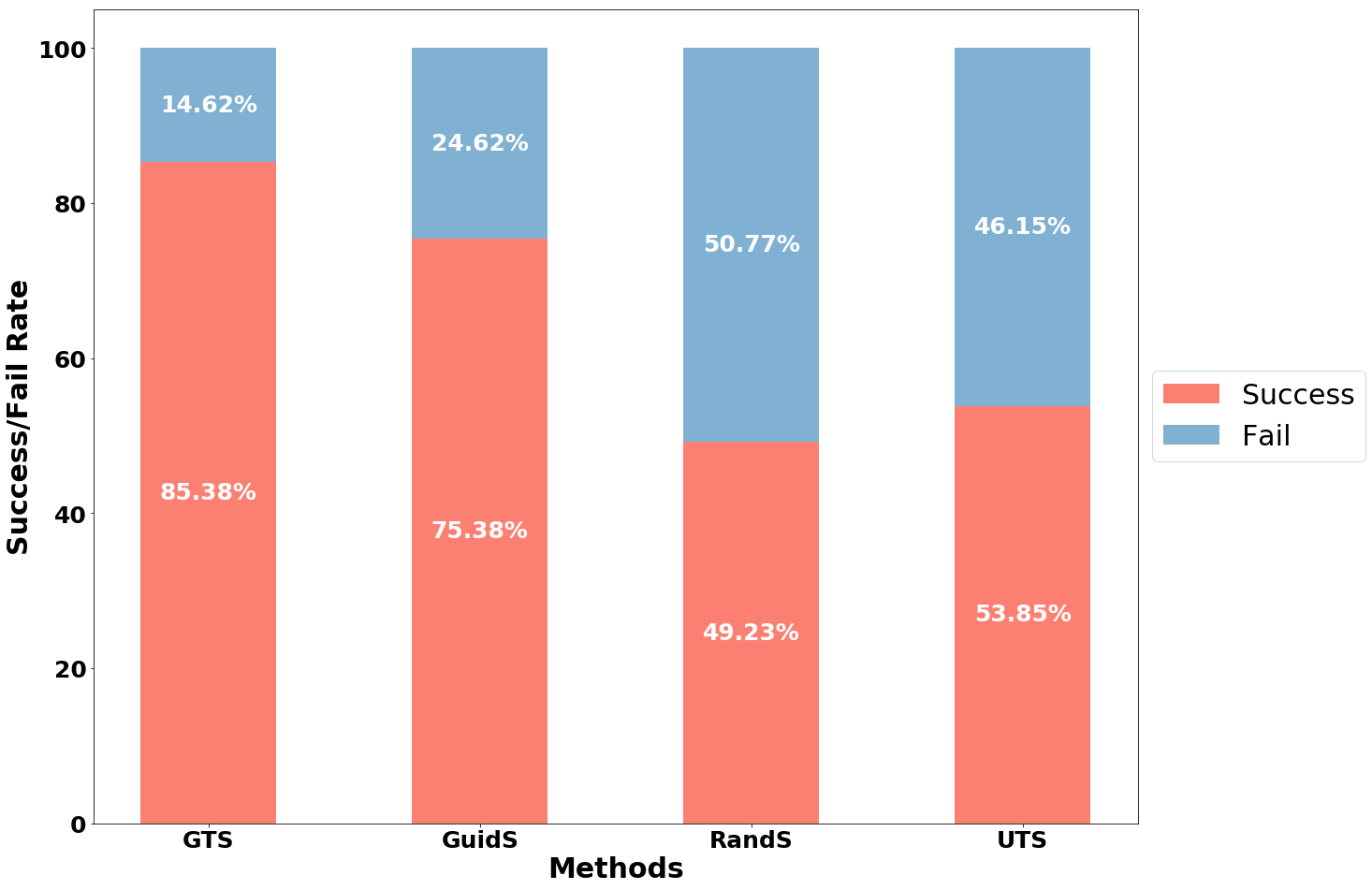}
        \caption{
        \label{fig:success-fail-all}}
    \end{subfigure}
    \vspace*{-2ex}
    \captionv{15}{success-fail rate}{The rate at which each method successfully found a solution with lower objective function value than that of CG solution. Each attempt is limited to 1000 seconds. \ref{fig:success-fail} The percentages of test cases that each method found a solution better than CG solution in at least 1 of their 5 attempts. \ref{fig:success-fail-all} The percentage of test cases that each method found a solution better than CG solution, averaged over all 5 of their attempts to solve the problem.}
\end{figure}

\section{Results}
At each attempt to solve a test scenario, each method is given 1000 seconds to search the solution space. Whenever a method finds a solution better than that of CG, the solution and its corresponding time stamp and the number of total solutions visited by this method are saved. We use these values to analyze the performance of each method. The best solution that is found in each attempt to solve the problem is used as the final solution of that attempt and is used for PTV metrics calculation. The average objective function value of final solutions in five attempts are used for the comparison of the performance of four methods with CG solution.
At first we compare the efficiency of the four methods of GTS, GuidS, RTS and RandS. Although the main purpose of these methods are to find a solution better than CG, there were some cases that none of these method could beat CG solution, either CG solution was very close to optimal, or there were several local optimums with wide search space which makes it very difficult to explore it efficiently, this is especially true for RTS and RandS methods. The percentages of total number of attempts that each method could successfully find a solution better than CG in at least one of five attempts and averaged over all attempts to solve the problem are presented in \autoref{fig:success-fail} and \ref{fig:success-fail-all}, respectively. Note that for this test the stopping criteria was 1000 seconds of computational time. As we expected, GTS and GuidS that are using the pre-trained DNN performed better than the other two methods. However, there are still cases that they are not able to find a better solution than CG. The maximum number of scenarios that all four methods were successfully find a solution with objective function value better than that of CG solution is 102 out of total 130 test cases (78.46\%). 

\begin{figure}[tb]
        \centering
        \includegraphics[width=.8\textwidth]{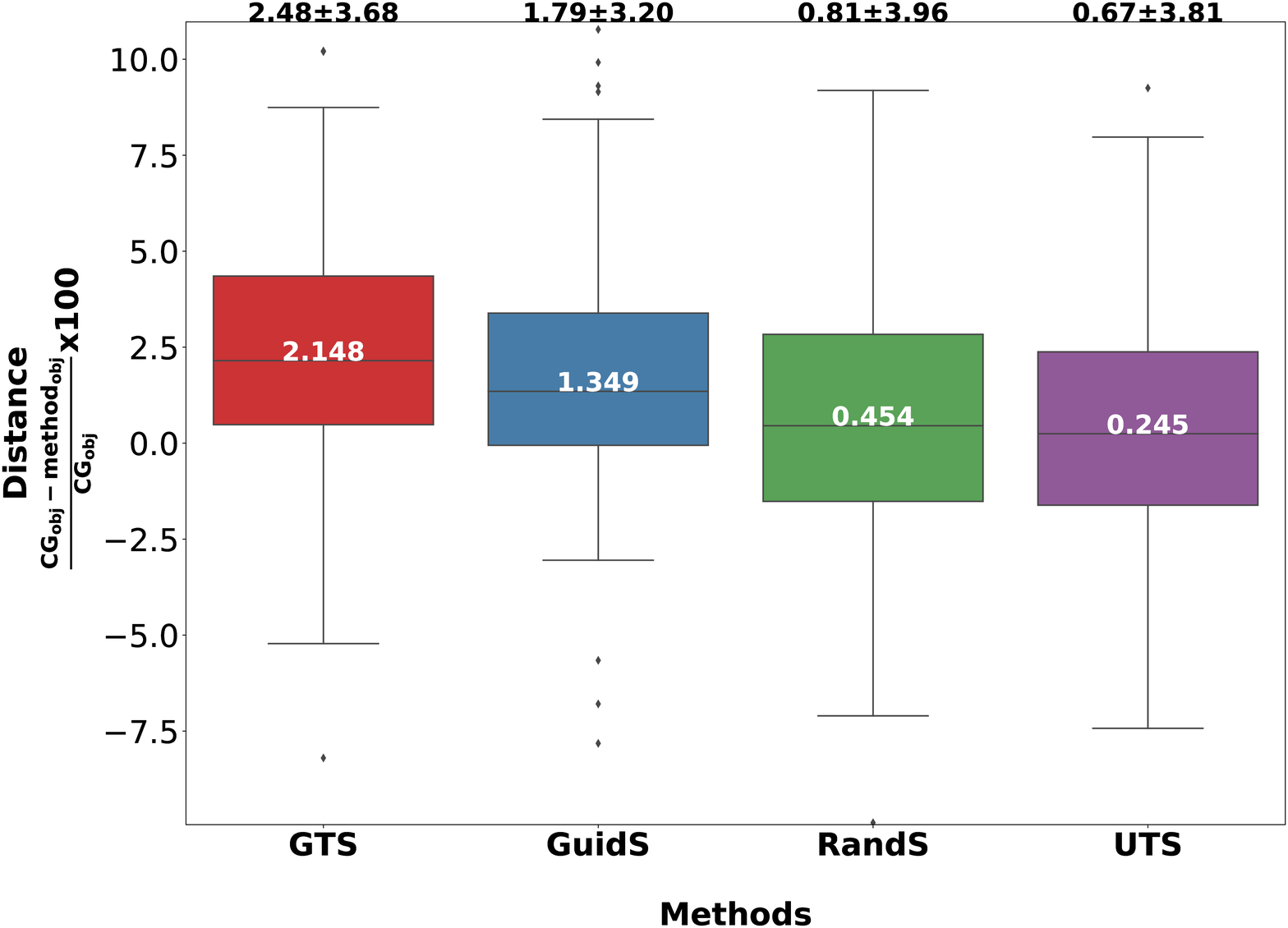}
        \vspace*{-3ex}
        \captionv{15}{Distance measure}{
        The distance measure of the average of the best objective function value found by each method compared to CG solution. The value written inside the box is the median and the value at the top of each box-plot is the mean $pm$ standard deviation of the Distance measure.
        \label{fig:avg-distance}}
\end{figure}

The domain of the objective function value varies for different test-case scenarios, therefore we introduce \textbf{Distance} measure as the normalized version of the objective values compared to CG solutions for further comparison. Distance measure is the difference between the objective function value of each method and CG, divided by CG objective value ($Distance=\frac{CG_{obj}-method_{obj}}{CG_obj}\times 100$). If a method finds a solution better than CG, $Distance$ measure will be positive, and for cases that a method was not successful to find a solution better than CG solution, this value will be negative. It means a method with largest $Distance$ measure found solutions with better qualities--with an objective value smaller than that of CG--and therefore more efficient in the limited time of 1000 seconds. \autoref{fig:avg-distance} shows the box plot of Distance measure for GTS, GuidS, RTS, and RandS methods. Based on this figure, the average $Distance$ measure for GTS method is 2.48, which is the highest compared to 1.79 for GuidS, 0.67 for RTS, and 0.81 for RandS.

\begin{figure}[tb]
    \begin{subfigure}{0.5\textwidth}
        \centering
    \includegraphics[clip,trim= 10mm 27mm 3mm 1mm, scale=.335]{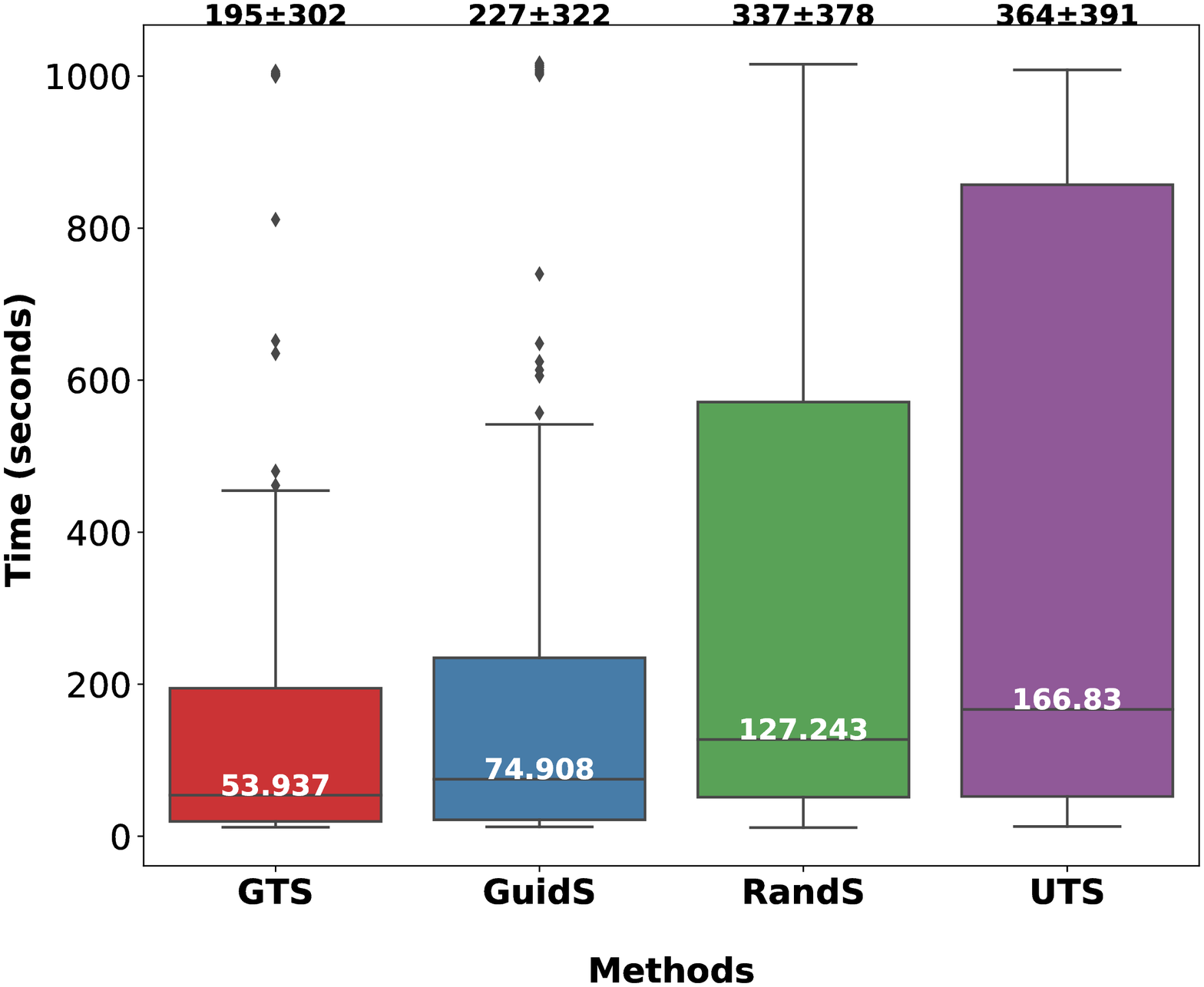}
        \captionv{6}{Best time}{
        The best time to beat CG solution.
        \label{fig:timelimit-Time}}
    \end{subfigure}
    \begin{subfigure}{.45\textwidth}
        \centering
        \includegraphics[clip,trim= 38.5mm 27mm 0mm 2mm,scale=0.335]{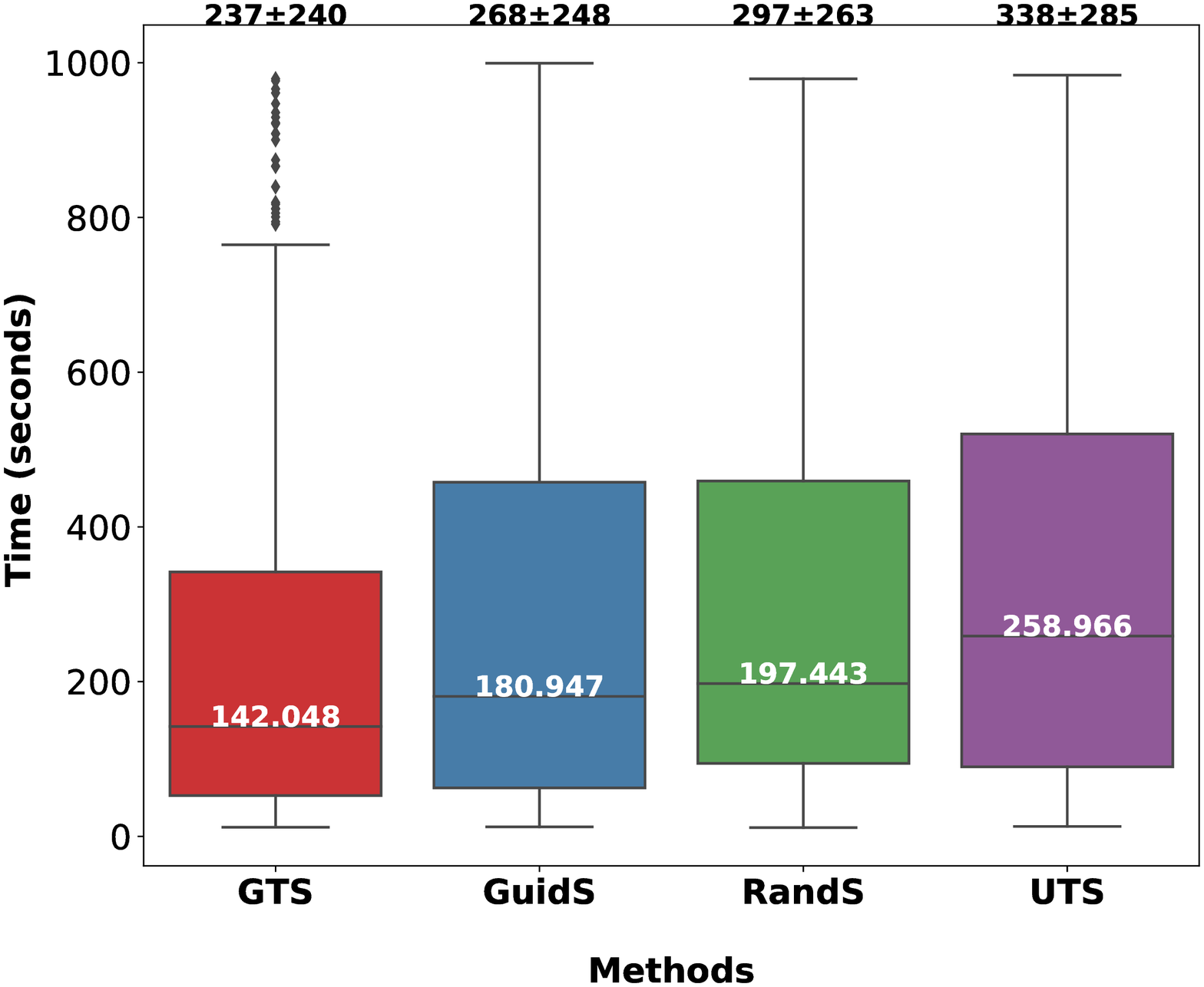}
        \captionv{5}{average time}{The average time to beat CG.
        \label{fig:timelimit-Time-obj}}
    \end{subfigure}
    \vspace*{-4ex}
    \captionv{15}{Time boxplot}{The computational time comparison between GTS, GuidS, RandS and RTS. The first time that each method beats a column generation solution. For this measure only successful scenarios are considered. Mean, standard deviation(at the top of each box as mean $pm$ standard deviation) and median (in the middle of the box) are calculated for each method.
    \label{fig:besttime}}
\end{figure}

For the computational time evaluation and comparison methods, only the 102 out of 130 test cases, where all four methods had successfully found a better solution than CG within the 1000 seconds time limit, were considered. The box-plot of the best time needed to beat CG solution is presented in \autoref{fig:timelimit-Time}. It represents the first time that a method finds a solution better than CG solution within 1000 seconds among all five attempts. The box-plot of the average time to beat CG among all attempts to solve each test-cases is provided in \autoref{fig:timelimit-Time-obj}. To measure the average time to beat CG, all test-cases and all attempts are considered. The fastest method considering this measure is GTS, the average time to beat CG for GTS method is 195 seconds in best case and 237 seconds in total cases. The second fast method is GuidS with the average time of 227 seconds for best and 268 seconds for total cases. RandS outperforms RTS with best time of 337 seconds compared to 364 seconds of RTS. Interestingly, the average total time of RTS and RandS is less than the average of best cases.

\begin{table}[tbh]
\begin{center}
\captionv{15}{One-tailed paired sample t-test for $\%99$ confidence interval }{One-tailed paired sample t-test to test the average objective function value and $Distance$ measure for every pairs of CG, GTS, GuidS, RTS, and RandS methods, with $\%99$ confidence intervals. All values in red have p-values greater than 0.01.)
\label{table:ttest-objValue}
}
    \begin{tabular}{lrrrr}
   \hline
          & \multicolumn{2}{c}{\textbf{Objective Value}} & \multicolumn{2}{c}{\textbf{Distance($\bf{\frac{CG-obj}{CG}}$)}} \\
    \hline
        \multicolumn{1}{l}{\textbf{Tested methods}} & \multicolumn{1}{r}{\textbf{t-statistic}} & \multicolumn{1}{r}{\textbf{p-value}} & \multicolumn{1}{r}{\textbf{t-statistic}} & \multicolumn{1}{r}{\textbf{ p-value  }} \\
    \hline
    \textbf{CG vs GTS} & 6.267 & 2.54E-09 & {7.683} & {1.72E-12} \\
    \textbf{CG vs GuidS} & 4.940 & 1.19E-06 & {6.393} & {1.37E-09} \\
    \textbf{CG vs RTS} & {\textbf{\color{red}0.885}} & {\textbf{\color{red}1.89E-01}} & {\textbf{\color{red}2.014}} & {\textbf{\color{red}2.30E-02}} \\
    \textbf{CG vs RandS} & {\textbf{\color{red}1.670}} & {\textbf{\color{red}4.87E-02}} & {\textbf{\color{red}2.340}} & {\textbf{\color{red}1.04E-02}} \\
    \textbf{RandS vs RTS} & {\textbf{\color{red}-1.496}} & {\textbf{\color{red}6.85E-02}} & {\textbf{\color{red}1.096}} & {\textbf{\color{red}1.38E-01}} \\
    \textbf{RandS vs GTS} & 6.826 & 1.53E-10 & -11.843 & 1.18E-22 \\
    \textbf{RandS vs GuidS} & 3.873 & 8.51E-05 & -4.839 & 1.83E-06 \\
    \textbf{RTS vs GTS} & 8.245 & 8.18E-14 & -15.271 & 5.25E-31 \\
    \textbf{RTS vs GuidS} & 5.257 & 2.95E-07 & -5.832 & 2.08E-08 \\
    \textbf{GuidS vs GTS} & 2.412 & 8.64E-03 & -3.945 & 6.52E-05 \\
    \hline
    \end{tabular}%
\end{center}
\end{table}

To study the statistical significant of GTS method compared to other four methods (GuidS, RTS, RandS, and CG), we use one-tailed paired sample t-test to compare the objective function values of each pair of methods, and Distance measure. The null hypothesis is that the average objective function and $Distance$ measure of all methods are the same. If we show the null hypothesis as $GTS=GuidS=RandS=RTS=CG=0$. The alternative hypothesis can be described as $GTS<GuidS<RandS<RTS<CG$ for the objective value parameters and $GTS>GuidS>RandS>RTS>CG$ for $Distance$ measure. Ten paired sample t-test are performed for objective function values and $Distance$ measure.These statistics are presented in \autoref{table:ttest-objValue}. The distribution of Objective value and $Distance$ measures are provided in appendix \autoref{appendix}.
As highlighted by red, all pairs of the three methods of CG, RTS, and RandS have p-values greater than 0.01, and are not significantly different, while the average $Distance$ and objective value measures of GuidS and GTS are significantly different. Based on these results GTS outperforms all other methods significantly while in the second position is GuidS as was expected.

\begin{table}[tb]
\begin{center}
\captionv{15}{PTV metrics}{Mean $\pm$ standard deviation for PTV Statistics, Paddick Conformity Index (${CI}_{Paddick}$), and High Dose Spillage ($R_{50}$) of methods: CG, GTS, GuidS, RTS, and RandS.
\label{table:PTV-metrics}
}
\resizebox{\textwidth}{!}{
\begin{tabular}{lcccccc}
\hline
\textbf{Method} & \textbf{PTV $\bf{D_{98}}$} & \textbf{PTV $\bf{D_{99}}$} & \textbf{PTV $\bf{D_{max}}$} & \textbf{PTV Homogeneity} & \textbf{$\bf{{CI}_{Paddick}}$} & \textbf{$\bf{R_{50}}$} \\
\hline
\textbf{RandS} & 0.977$\pm$0.012 & 0.960$\pm$0.019 & 1.071$\pm$0.040 & 0.089$\pm$0.046 & 0.867$\pm$0.070 & 4.673$\pm$1.022 \\
\textbf{RTS} & 0.977$\pm$0.011 & 0.959$\pm$0.020 & 1.070$\pm$0.039 & 0.088$\pm$0.045 & 0.863$\pm$0.085 & 4.714$\pm$1.312 \\
\textbf{GTS} & 0.977$\pm$0.011 & 0.960$\pm$0.019 & 1.070$\pm$0.040 & 0.089$\pm$0.045 & 0.874$\pm$0.061 & 4.569$\pm$0.994 \\
\textbf{GuidS} & 0.976$\pm$0.012 & 0.960$\pm$0.019 & 1.071$\pm$0.040 & 0.089$\pm$0.046 & 0.874$\pm$0.068 & 4.487$\pm$0.948 \\
\textbf{CG} & 0.977$\pm$0.011 & 0.961$\pm$0.020 & 1.072$\pm$0.041 & 0.090$\pm$0.046 & 0.884$\pm$0.059 & 4.478$\pm$0.963 \\
\hline
    \end{tabular}%
    }
\end{center}
\end{table}

PTV statistics, Paddick Conformity Index(${CI}_{Paddick}$) and dose spillage($R_{50}$) of plans generated by CG, GTS, GuidS, RTS, and RandS are presented in \autoref{table:PTV-metrics}. Note that PTV $D_2$ is used to measure PTV $D_{max}$, as recommended by the ICRU Report 83 \citep{Hodapp2012TheIMRT}. The plans generated by all methods have very similar PTV coverage. CG plans have the highest ${CI}_{Paddick}$ followed by GTS and GuidS plans. While CG and GuidS plans have the lowest dose spillage value  followed by GTS. 

\begin{table}[bth]
\begin{center}
\captionv{16}{Mean and Max Dose}{The average and maximum fractional dose received by each structure in plans generated by GTS, GuidS, RTS, RandS, and, CG methods, where prescription dose is set to 1. 
\label{table:maxmeandose}
\vspace*{-2ex}
}
\resizebox{1.\textwidth}{!}{
\begin{tabular}{cllllll}
      &       & \multicolumn{5}{c}{\textbf{Methods}} \\
\hline
      & \textbf{Structures} & \textbf{GTS} & \textbf{GuidS} & \textbf{RTS} & \textbf{RandS} & \textbf{CG} \\
\hline
\multirow{6}{*}{\rotatebox{90}{\textbf{Mean Dose}}} 
      & \textbf{PTV} & 1.039  $\pm$ 0.025 & 1.039  $\pm$ 0.025 & 1.038  $\pm$ 0.024 & 1.039  $\pm$ 0.025 & 1.040  $\pm$ 0.025 \\
      & \textbf{Body} & \textbf{0.037  $\pm$ 0.012} & \textbf{0.037  $\pm$ 0.012} & \textbf{0.037  $\pm$ 0.012} & \textbf{0.037  $\pm$ 0.012} & 0.038  $\pm$ 0.013 \\
      & \textbf{Bladder} & 0.207  $\pm$ 0.125 & 0.207  $\pm$ 0.122 & 0.207  $\pm$ 0.126 & 0.206  $\pm$ 0.122 & \textbf{0.204  $\pm$ 0.116}  \\
      & \textbf{Rectum} & \textbf{0.317  $\pm$ 0.109} & 0.321  $\pm$ 0.111 & 0.319  $\pm$ 0.110 & 0.322  $\pm$ 0.115 & 0.334  $\pm$ 0.116  \\
      & \textbf{L-femoral} & 0.213  $\pm$ 0.105 & \textbf{0.201  $\pm$ 0.103} & 0.217  $\pm$ 0.115 & 0.212  $\pm$ 0.111 & 0.222  $\pm$ 0.112 \\
      & \textbf{R-femoral} & \textbf{0.214  $\pm$ 0.101} & 0.221  $\pm$ 0.110 & 0.227  $\pm$ 0.124 & 0.224  $\pm$ 0.124 & 0.217  $\pm$ 0.109 \\
\hline 
\multirow{6}{*}{\rotatebox{90}{\textbf{Max Dose}}} & \textbf{PTV} & 1.113  $\pm$ 0.055 & 1.113  $\pm$ 0.055 & 1.113  $\pm$ 0.055 & 1.114  $\pm$ 0.057 & 1.116  $\pm$ 0.058 \\
      & \textbf{Body} & 1.190  $\pm$ 0.131 & 1.195  $\pm$ 0.148 & 1.200  $\pm$ 0.147 & 1.199  $\pm$ 0.143 & \textbf{1.173  $\pm$ 0.130} \\
      & \textbf{Bladder} & \textbf{1.094  $\pm$ 0.046} & 1.094  $\pm$ 0.048 & 1.096  $\pm$ 0.050 & \textbf{1.094  $\pm$ 0.045} & 1.095  $\pm$ 0.048  \\
      & \textbf{Rectum} & 1.072  $\pm$ 0.045 & 1.071  $\pm$ 0.044 & 1.073  $\pm$ 0.045 & 1.074  $\pm$ 0.048 & \textbf{1.071  $\pm$ 0.040} \\
      & \textbf{L-femoral} & 0.609  $\pm$ 0.193 & \textbf{0.596  $\pm$ 0.209} & 0.619  $\pm$ 0.215 & 0.609  $\pm$ 0.220 & 0.613  $\pm$ 0.193  \\
      & \textbf{R-femoral} & 0.639  $\pm$ 0.242 & 0.650  $\pm$ 0.249 & 0.625  $\pm$ 0.236 & 0.628  $\pm$ 0.244 & \textbf{0.617  $\pm$ 0.245} \\
\hline\\
\end{tabular}%
}
\end{center}
\end{table}

\begin{figure}[htb]
    \centering
    \includegraphics[width=1\textwidth]{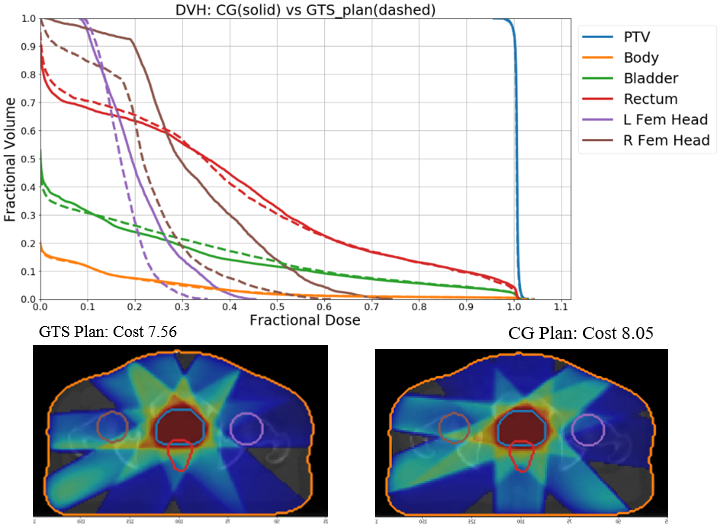}
    \captionv{15}{CG vs GTS}{ GTS generated plan (dashed) vs CG generated plan (solid).
    \label{fig:cgvsGTS}}
\end{figure}

The average and maximum dose received by each structure are provided in \autoref{table:maxmeandose}, these values reflect the fractional dose in plans generated by each method with the assumption that the prescription dose is one-- e.g. if the prescription dose is 70 Gy, the average dose of 0.207 in the table means 14.47Gy ($0.207\times 70$) in the prescribed plan. The minimum values in each row are shown as bold numbers for easier interpretation of the table. On average plans generated by GTS have lower mean dose to OARs compared to other methods, while plans generated by CG have the lowest maximum dose to OARs. GTS plans spare rectum and right femoral head better than other methods. Although the average fractional dose to bladder by by GTS plans (0.207) is more than CG plans (0.204), GTS plans have lower maximum fractional dose to bladder (1.094) compared to CG (with maximum of 1.095). GuidS plans have minimum average fractional dose to left-femoral head (0.201) with considerable difference compared to the second best (0.212) of RandS plans. As an example, \autoref{fig:cgvsGTS} shows the dose-volume and dose-wash of plans generated by GTS and CG for one test-case scenario.

\section{Discussion}
In this research, we propose an efficient beam orientation optimization framework capable of finding a improved solution over CG, in a similar amount time, by utilizing a reinforcement learning structure involving a supervised learning network, DNN, to guide Monte Carlo tree search to explore the beam orientation selection decision space.

Although CG is a powerful optimization tool, it is a greedy approach that not only is computationally expensive and time consuming, but it also may get stuck in a local optimum. This is particularly true for highly non-linear optimization problems with lots of local optima, such as BOO. In this work, we tried 4 different approaches: 1) Guided Tree Search (GTS), 2) Guided Search (GuidS), 3) Random Tree Seach (RTS), and 4) Random Search (RandS). It is shown that although the quality of solutions using RandS, RTS and CG were not significantly different, in $50\%$ of test-cases both RandS and RTS, which have no knowledge of the problem at the beginning of the search, can find solutions better than CG. This shows the high potential of improving the solution found by CG. 

We saw that GTS and GuidS, both performs better than other methods, which is expected because both of these methods are using a prior knowledge (trained DNN) to explore the solution space. GTS even outperforms GuidS on average, since GTS is a combination of GuidS and RTS, means adding a search method to GuidS can improve the quality of the solution. But considering the insignificant difference in the performance of RTS and RandS, adding any search method to GuidS may results in better solutions and it may not be directly related to RTS. This issue will be studied in future researches. The poor performance of RTS may also suggest that using uniform tree search is too slow to converge to the optimal selection of beams.

Although GTS performs better than others to find solutions with better objective function values, but the dose spillage metrics, and specifically the average dose received by bladder in GTS plans can be improved further. Considering the success of GTS in reducing the objective function and its potential for further improvement, we will continue exploring new methods and techniques to upgrade the quality of treatment planning with the help of artificial intelligence.

We should note that CG is a greedy and deterministic algorithm, therefor using CG on the same problem always results in the same solution. This is of completely different nature from our search methods and it may not be fair to compare its performance with the four search procedures, which, given infinite time and resources, can act as brute forced approach and guarantee finding the optimal solution. However, our main goal is to find the best possible solution for BOO problem, and in this work we try to see which search algorithms can find a better solutions and faster. Hence we expect to see search algorithms outperform greedy algorithms. The results showed us that the objective function value of GTS and GuidS CG, RTS and RandS perform similarly.  Even though, plans generated by DNN solutions may not be superior to CG, but it can mimic the CG algorithm very efficiently \citep{SadeghnejadBarkousaraie2019ATherapy} and is a very successful tool to explore the search space as shown by GuidS and GTS, especially when compared to CG , which can easily exhaust the computational resources and is very slow to find one solution for a problems. The good performance of CG compared to RandS and RTS shows how powerful the CG method can be to find a solution, and the success of using DNN to explore the decision space represents the proper knowledge that can be achieved by learning from CG. 

Finally, GTS is a problem specific search method that needs to be applied on each test-cases separately. To use the knowledge that we can get from the GTS performance, more advanced reinforcement algorithms can be trained to create a single general knowledge-based method that is not only very powerful to find the best possible solutions, but also very fast for doing so. The advanced reinforcement learning method then can be easily applied on more sophisticated and challenging problems such as Proton and $4\pi$ radiation therapy. For future studies, we are working to develop a smart, fast and powerful tool to be applied in these problems. 


\section{Conclusion}
In this study, we proposed a method combined of two main components. First, a supervised deep neural network (DNN) to learn column generation (CG) decision-making process, and predict the fitness values of all candidate beams, beam with maximum fitness value will be chosen to add to the solution. CG, although powerful, is a heuristic, greedy algorithm that cannot guarantee the global optimality of the final solution, and leaves room for improvement. A Monte Carlo guided tree search (GTS) is proposed to see if finding a solution with better objective function in reasonable amount of time is feasible. After the DNN is trained, it is used to generate the beams fitness values for nodes in the decision tree, where each node represents a set of selected beams. Fitness values in each node are normalized and used as probability mass function to help deciding decision tree extension. Later probability distribution of beam selection will be modified by reward function, which is based on the solution of FMO problem FMO for every five selected beams. GTS continues to explore the decision tree for 1000 seconds. Along with GTS, three other approaches are also tested, GuidS which is also using DNN to select beams iteratively, but it does not update the probability distribution of beams during the search process. RTS which is a simple tree search algorithm, starts by randomly sampled from a uniform distribution of beam orientations for each node and continues to update beams probability distribution based on the tree search approach presented for GTS. And finally RandS which is randomly select beams, the most trivial and simple approach. 

\clearpage

\section{Appendix\label{appendix}}
\begin{figure}[htb]
    \begin{subfigure}{\textwidth}
    \centering
    \vspace{12pt}
    \includegraphics[clip,trim=0mm 0mm 103mm 0mm,width=\textwidth]{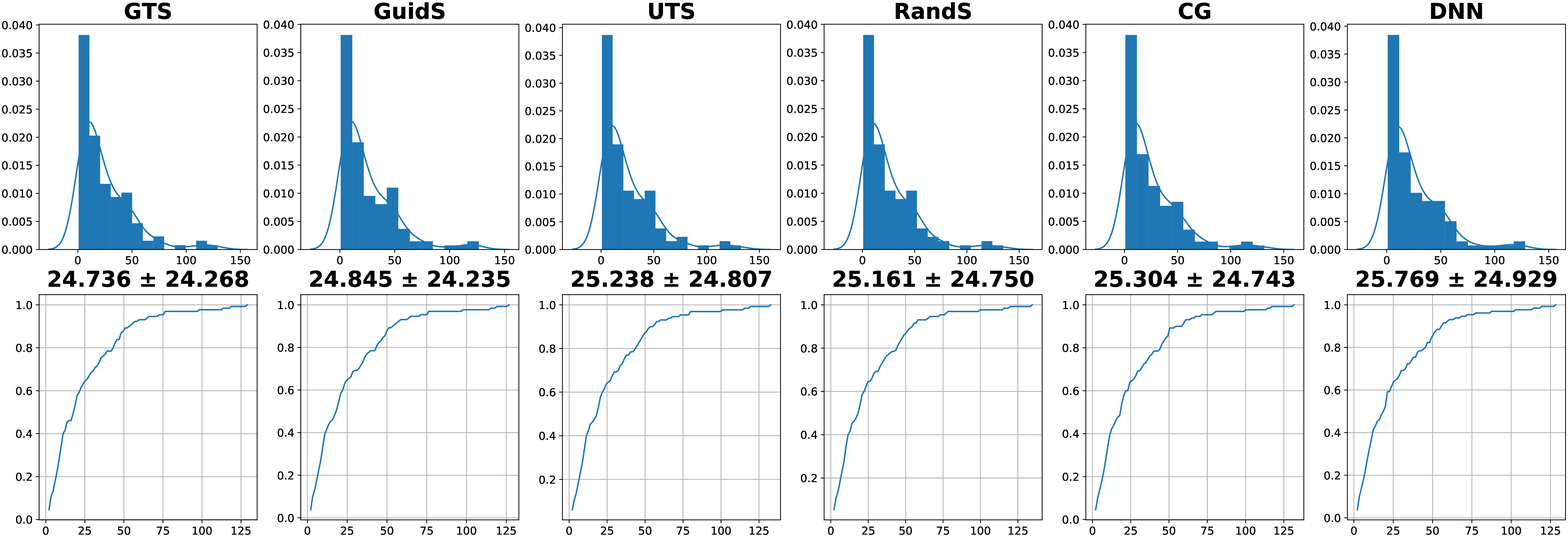}
    \captionv{15}{Objective values Distribution}{The distribution of objective values using GTS, GuidS, RTS, RandS, and CG  methods.
    \label{fig:objVal-distribution}}
    \end{subfigure}
    \begin{subfigure}{ \textwidth}
    \centering
    \includegraphics[clip, trim=0mm 0mm 120mm 0mm,width=\textwidth ]{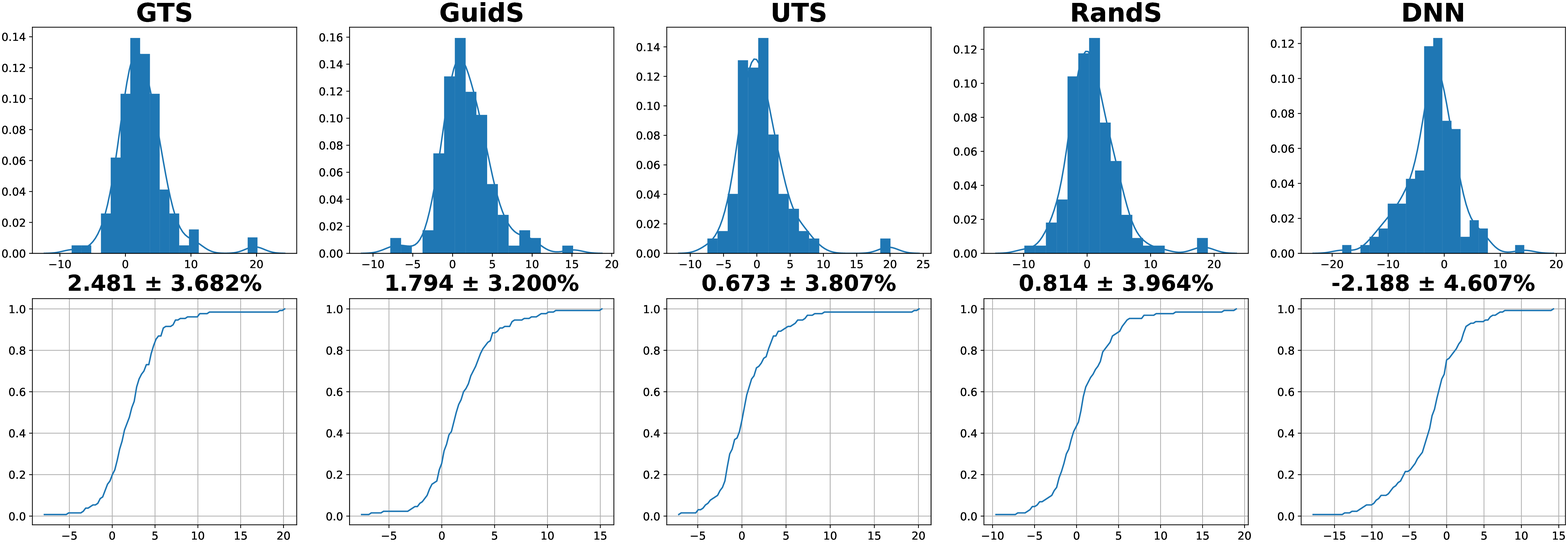}
    \captionv{15}{Distance Distribution}{Distance distribution of methods GTS, GuidS, RTS, and RandS.
    \label{fig:distance-distribution}}
    \end{subfigure}
    \captionv{15}{Distribution}{The distribution of objective values and distance to CG objective values\label{fig:distribution}}
\end{figure}

\vspace*{-10mm}
Although statistically with 130 number of test cases we can assume that approximately our metrics follow normal distribution, but based on the graphs of \autoref{fig:objVal-distribution}, this assumption may not be practical. Because of this, we introduce $Distance$ measures to normalize our metrics. The  probability distribution and cumulative probability mass function of $Distance$ measure  are presented in \autoref{fig:distance-distribution}. By this graph we can verify that the this measure approximately normally distributed with similar standard deviation.

\section*{References}


\vspace*{-10mm}
\begin{thebibliography}{49}
\providecommand{\natexlab}[1]{#1}
\providecommand{\url}[1]{\texttt{#1}}
\expandafter\ifx\csname urlstyle\endcsname\relax
  \providecommand{\doi}[1]{doi: #1}\else
  \providecommand{\doi}{doi: \begingroup \urlstyle{rm}\Url}\fi

    \bibitem[Aleman et~al.(2008)Aleman, Kumar, Ahuja, Romeijn, and
      Dempsey]{Aleman2008}
    Dionne~M. Aleman, Arvind Kumar, Ravindra~K. Ahuja, H.~Edwin Romeijn, and
      James~F. Dempsey.
    \newblock {Neighborhood search approaches to beam orientation optimization in
      intensity modulated radiation therapy treatment planning}.
    \newblock \emph{Journal of Global Optimization}, 42\penalty0 (4):\penalty0
      587--607, 12 2008.
    \newblock ISSN 09255001.
    \newblock \doi{10.1007/s10898-008-9286-x}.
    
    \bibitem[Amit et~al.(2015)Amit, Purdie, Levinshtein, Hope, Lindsay, Marshall,
      Jaffray, and Pekar]{Amit2015}
    Guy Amit, Thomas~G. Purdie, Alex Levinshtein, Andrew~J. Hope, Patricia Lindsay,
      Andrea Marshall, David~A. Jaffray, and Vladimir Pekar.
    \newblock {Automatic learning-based beam angle selection for thoracic IMRT}.
    \newblock \emph{Medical Physics}, 42\penalty0 (4), 4 2015.
    \newblock ISSN 00942405.
    \newblock \doi{10.1118/1.4908000}.
    
    \bibitem[Azizi-Sultan(2006)]{Azizi-Sultan2006}
    Ahmad~Saher Azizi-Sultan.
    \newblock \emph{{Optimization of Beam Orientations in Intensity Modulated
      Radiation Therapy}}.
    \newblock PhD thesis, Technische Universit{\"{a}}t Kaiserslautern, 2006.
    
    \bibitem[Bangert and Oelfke(2010)]{Bangert2010}
    Mark Bangert and Uwe Oelfke.
    \newblock {Spherical cluster analysis for beam angle optimization in
      intensity-modulated radiation therapy treatment planning}.
    \newblock \emph{Physics in Medicine and Biology}, 55\penalty0 (19):\penalty0
      6023--6037, 10 2010.
    \newblock ISSN 00319155.
    \newblock \doi{10.1088/0031-9155/55/19/025}.
    
    \bibitem[Bedford et~al.(2019)Bedford, Ziegenhein, Nill, and
      Oelfke]{Bedford2019}
    James~L. Bedford, Peter Ziegenhein, Simeon Nill, and Uwe Oelfke.
    \newblock {Beam selection for stereotactic ablative radiotherapy using
      Cyberknife with multileaf collimation}.
    \newblock \emph{Medical Engineering and Physics}, 64:\penalty0 28--36, 2 2019.
    \newblock ISSN 18734030.
    \newblock \doi{10.1016/j.medengphy.2018.12.011}.
    \newblock URL
      \url{https://www.sciencedirect.com/science/article/pii/S1350453318301802}.
    
    \bibitem[Bortfeld and Schlegel(1993)]{Bortfeld1993OptimizationConsiderations}
    T.~Bortfeld and W.~Schlegel.
    \newblock {Optimization of beam orientations in radiation therapy: Some
      theoretical considerations}.
    \newblock \emph{Physics in Medicine and Biology}, 38\penalty0 (2):\penalty0
      291--304, 1993.
    \newblock ISSN 00319155.
    \newblock \doi{10.1088/0031-9155/38/2/006}.
    
    \bibitem[Breedveld et~al.(2009)Breedveld, Storchi, and Heijmen]{Breedveld2009}
    Sebastiaan Breedveld, Pascal~R.M. Storchi, and Ben~J.M. Heijmen.
    \newblock {The equivalence of multi-criteria methods for radiotherapy plan
      optimization.}
    \newblock \emph{Physics in medicine and biology}, 54\penalty0 (23):\penalty0
      7199--7209, 2009.
    \newblock ISSN 13616560.
    \newblock \doi{10.1088/0031-9155/54/23/011}.
    
    \bibitem[Breedveld et~al.(2012)Breedveld, Storchi, Voet, and
      Heijmen]{Breedveld2012}
    Sebastiaan Breedveld, Pascal~R.M. Storchi, Peter~W.J. Voet, and Ben~J.M.
      Heijmen.
    \newblock {ICycle: Integrated, multicriterial beam angle, and profile
      optimization for generation of coplanar and noncoplanar IMRT plans}.
    \newblock \emph{Medical Physics}, 39\penalty0 (2):\penalty0 951--963, 2012.
    \newblock ISSN 00942405.
    \newblock \doi{10.1118/1.3676689}.
    
    \bibitem[Browne et~al.(2012)Browne, Powley, Whitehouse, Lucas, Cowling,
      Rohlfshagen, Tavener, Perez, Samothrakis, and Colton]{Browne2012}
    Cameron~B. Browne, Edward Powley, Daniel Whitehouse, Simon~M. Lucas, Peter~I.
      Cowling, Philipp Rohlfshagen, Stephen Tavener, Diego Perez, Spyridon
      Samothrakis, and Simon Colton.
    \newblock {A survey of Monte Carlo tree search methods}, 3 2012.
    \newblock ISSN 1943068X.
    
    \bibitem[Cabrera~G. et~al.(2018)Cabrera~G., Ehrgott, Mason, and
      Raith]{CabreraG.2018}
    Guillermo Cabrera~G., Matthias Ehrgott, Andrew~J. Mason, and Andrea Raith.
    \newblock {A matheuristic approach to solve the multiobjective beam angle
      optimization problem in intensity-modulated radiation therapy}.
    \newblock \emph{International Transactions in Operational Research},
      25\penalty0 (1):\penalty0 243--268, 1 2018.
    \newblock ISSN 14753995.
    \newblock \doi{10.1111/itor.12241}.
    
    \bibitem[Cabrera-Guerrero et~al.(2018{\natexlab{a}})Cabrera-Guerrero, Lagos,
      Cabrera, Johnson, Rubio, and
      Paredes]{Cabrera-Guerrero2018ComparingRadiotherapy}
    Guillermo Cabrera-Guerrero, Carolina Lagos, Enrique Cabrera, Franklin Johnson,
      Jose~M. Rubio, and Fernando Paredes.
    \newblock {Comparing Local Search Algorithms for the Beam Angles Selection in
      Radiotherapy}.
    \newblock \emph{IEEE Access}, 6:\penalty0 23701--23710, 4 2018{\natexlab{a}}.
    \newblock ISSN 21693536.
    \newblock \doi{10.1109/ACCESS.2018.2830646}.
    
    \bibitem[Cabrera-Guerrero et~al.(2018{\natexlab{b}})Cabrera-Guerrero, Mason,
      Raith, and Ehrgott]{Cabrera-Guerrero2018}
    Guillermo Cabrera-Guerrero, Andrew~J. Mason, Andrea Raith, and Matthias
      Ehrgott.
    \newblock {Pareto local search algorithms for the multi-objective beam angle
      optimisation problem}.
    \newblock \emph{Journal of Heuristics}, 24\penalty0 (2):\penalty0 205--238, 4
      2018{\natexlab{b}}.
    \newblock ISSN 15729397.
    \newblock \doi{10.1007/s10732-018-9365-1}.
    
    \bibitem[Chambolle and Pock(2010)]{Chambolle2010}
    Antonin Chambolle and Thomas Pock.
    \newblock {A first-order primal-dual algorithm for convex problems with
      applications to imaging}, 2010.
    \newblock URL \url{https://hal.archives-ouvertes.fr/hal-00490826}.
    
    \bibitem[Craft and Monz(2010)]{Craft2010}
    David Craft and Michael Monz.
    \newblock {Simultaneous navigation of multiple Pareto surfaces, with an
      application to multicriteria IMRT planning with multiple beam angle
      configurations}.
    \newblock \emph{Medical Physics}, 37\penalty0 (2):\penalty0 736--741, 2010.
    \newblock ISSN 00942405.
    \newblock \doi{10.1118/1.3292636}.
    
    \bibitem[Djajaputra et~al.(2003)Djajaputra, Wu, Wu, and Mohan]{Djajaputra2003}
    David Djajaputra, Qiuwen Wu, Yan Wu, and Radhe Mohan.
    \newblock {Algorithm and performance of a clinical IMRT beam-angle optimization
      system}, 2003.
    \newblock ISSN 00319155.
    
    \bibitem[Dong et~al.(2013)Dong, Lee, Ruan, Long, Romeijn, Yang, Low, Kupelian,
      and Sheng]{Dong2013}
    Peng Dong, Percy Lee, Dan Ruan, Troy Long, Edwin Romeijn, Yingli Yang, Daniel
      Low, Patrick Kupelian, and Ke~Sheng.
    \newblock {4{$\pi$} non-coplanar liver SBRT: A novel delivery technique}.
    \newblock \emph{International Journal of Radiation Oncology Biology Physics},
      85\penalty0 (5):\penalty0 1360--1366, 4 2013.
    \newblock ISSN 03603016.
    \newblock \doi{10.1016/j.ijrobp.2012.09.028}.
    
    \bibitem[Gu et~al.(2018)Gu, O'Connor, Nguyen, Yu, Ruan, Dong, and
      Sheng]{Gu2018}
    Wenbo Gu, Daniel O'Connor, Dan Nguyen, Victoria~Y. Yu, Dan Ruan, Lei Dong, and
      Ke~Sheng.
    \newblock {Integrated beam orientation and scanning-spot optimization in
      intensity-modulated proton therapy for brain and unilateral head and neck
      tumors}.
    \newblock \emph{Medical Physics}, 45\penalty0 (4):\penalty0 1338--1350, 4 2018.
    \newblock ISSN 00942405.
    \newblock \doi{10.1002/mp.12788}.
    
    \bibitem[Gu et~al.(2019)Gu, Neph, Ruan, Zou, Dong, and Sheng]{Gu2019}
    Wenbo Gu, Ryan Neph, Dan Ruan, Wei Zou, Lei Dong, and Ke~Sheng.
    \newblock {Robust Beam Orientation Optimization for Intensity Modulated
      Proton Therapy}.
    \newblock \emph{Medical Physics}, 46\penalty0 (8):\penalty0 mp.13641, 6 2019.
    \newblock ISSN 0094-2405.
    \newblock \doi{10.1002/mp.13641}.
    \newblock URL \url{https://onlinelibrary.wiley.com/doi/abs/10.1002/mp.13641}.
    
    \bibitem[Hodapp(2012)]{Hodapp2012TheIMRT}
    N.~Hodapp.
    \newblock {The ICRU Report No. 83: Prescribing, recording and reporting
      photon-beam intensity-modulated radiation therapy (IMRT)}, 1 2012.
    \newblock ISSN 01797158.
    
    \bibitem[Karush(2014)]{Karush2014}
    William Karush.
    \newblock \emph{Minima of Functions of Several Variables with Inequalities as
      Side Conditions}, pages 217--245.
    \newblock Springer Basel, Basel, 2014.
    \newblock ISBN 978-3-0348-0439-4.
    \newblock \doi{10.1007/978-3-0348-0439-4_10}.
    \newblock URL \url{https://doi.org/10.1007/978-3-0348-0439-4_10}.
    
    \bibitem[Kuhn and Tucker(1951)]{kuhn1951nonlinear}
    Harold~W Kuhn and Albert~W Tucker.
    \newblock Nonlinear programming, in (j. neyman, ed.) proceedings of the second
      berkeley symposium on mathematical statistics and probability, 1951.
    
    \bibitem[Li et~al.(2004)Li, Yao, and Yao]{Li2004AutomaticAlgorithm}
    Yongjie Li, Jonathan Yao, and Dezhong Yao.
    \newblock {Automatic beam angle selection in IMRT planning using genetic
      algorithm}.
    \newblock \emph{Physics in Medicine and Biology}, 49\penalty0 (10):\penalty0
      1915--1932, 5 2004.
    \newblock ISSN 00319155.
    \newblock \doi{10.1088/0031-9155/49/10/007}.
    
    \bibitem[Li et~al.(2005)Li, Yao, Yao, and Chen]{Li2005}
    Yongjie Li, Dezhong Yao, Jonathan Yao, and Wufan Chen.
    \newblock {A particle swarm optimization algorithm for beam angle selection in
      intensity-modulated radiotherapy planning}.
    \newblock \emph{Physics in Medicine and Biology}, 50\penalty0 (15):\penalty0
      3491--3514, 8 2005.
    \newblock ISSN 00319155.
    \newblock \doi{10.1088/0031-9155/50/15/002}.
    
    \bibitem[Lim et~al.(2009)Lim, Holder, and Reese]{Lim2009APlanning}
    Gino Lim, Allen Holder, and Josh Reese.
    \newblock {A clustering approach for optimizing beam angles in IMRT planning}.
    \newblock \emph{Mathematical Sciences Technical Reports (MSTR)}, 8 2009.
    \newblock URL \url{https://scholar.rose-hulman.edu/math_mstr/14}.
    
    \bibitem[Lim et~al.(2008)Lim, Choi, and Mohan]{Lim2008}
    Gino~J. Lim, Jaewon Choi, and Radhe Mohan.
    \newblock {Iterative solution methods for beam angle and fluence map
      optimization in intensity modulated radiation therapy planning}.
    \newblock \emph{OR Spectrum}, 30\penalty0 (2):\penalty0 289--309, 4 2008.
    \newblock ISSN 01716468.
    \newblock \doi{10.1007/s00291-007-0096-1}.
    
    \bibitem[Liu et~al.(2017)Liu, Dong, and Xing]{Liu2017a}
    Hongcheng Liu, Peng Dong, and Lei Xing.
    \newblock {A new sparse optimization scheme for simultaneous beam angle and
      fluence map optimization in radiotherapy planning}.
    \newblock \emph{Physics in Medicine and Biology}, 62\penalty0 (16):\penalty0
      6428--6445, 7 2017.
    \newblock ISSN 13616560.
    \newblock \doi{10.1088/1361-6560/aa75c0}.
    
    \bibitem[Llacer et~al.(2009)Llacer, Li, Agazaryan, Promberger, and
      Solberg]{Llacer2009}
    Jorge Llacer, Sicong Li, Nzhde Agazaryan, Claus Promberger, and Timothy~D.
      Solberg.
    \newblock {Non-coplanar automatic beam orientation selection in cranial IMRT: A
      practical methodology}.
    \newblock \emph{Physics in Medicine and Biology}, 54\penalty0 (5):\penalty0
      1337--1368, 2009.
    \newblock ISSN 00319155.
    \newblock \doi{10.1088/0031-9155/54/5/016}.
    
    \bibitem[Nguyen et~al.(2016)Nguyen, Thomas, Cao, O'Connor, Lamb, and
      Sheng]{Nguyen2016}
    Dan Nguyen, David Thomas, Minsong Cao, Daniel O'Connor, James Lamb, and
      Ke~Sheng.
    \newblock {Computerized triplet beam orientation optimization for MRI-guided
      Co-60 radiotherapy}.
    \newblock \emph{Medical Physics}, 43\penalty0 (10):\penalty0 5667--5675, 10
      2016.
    \newblock ISSN 00942405.
    \newblock \doi{10.1118/1.4963212}.
    
    \bibitem[O'Connor et~al.(2018)O'Connor, Yu, Nguyen, Ruan, and
      Sheng]{OConnor2018}
    Daniel O'Connor, Victoria Yu, Dan Nguyen, Dan Ruan, and Ke~Sheng.
    \newblock {Fraction-variant beam orientation optimization for non-coplanar
      IMRT}.
    \newblock \emph{Physics in Medicine and Biology}, 63\penalty0 (4), 2 2018.
    \newblock ISSN 13616560.
    \newblock \doi{10.1088/1361-6560/aaa94f}.
    
    \bibitem[Oelfke and Bortfeld(2001)]{Oelfke2001}
    U.~Oelfke and T.~Bortfeld.
    \newblock {Inverse planning for photon and proton beams}.
    \newblock \emph{Medical Dosimetry}, 26\penalty0 (2):\penalty0 113--124, 2001.
    \newblock URL
      \url{http://www.sciencedirect.com/science/article/pii/S0958394701000577}.
    
    \bibitem[Paddick(2000)]{paddick2000simple}
    Ian Paddick.
    \newblock A simple scoring ratio to index the conformity of radiosurgical
      treatment plans.
    \newblock \emph{Journal of neurosurgery}, 93\penalty0 (supplement\_3):\penalty0
      219--222, 2000.
    
    \bibitem[Potrebko et~al.(2008)Potrebko, McCurdy, Butler, and
      El-Gubtan]{Potrebko2008}
    Peter~S. Potrebko, Boyd~M.C. McCurdy, James~B. Butler, and Adel~S. El-Gubtan.
    \newblock {Improving intensity-modulated radiation therapy using the anatomic
      beam orientation optimization algorithm}.
    \newblock \emph{Medical Physics}, 35\penalty0 (5):\penalty0 2170--2179, 2008.
    \newblock ISSN 00942405.
    \newblock \doi{10.1118/1.2905026}.
    
    \bibitem[Pugachev et~al.(2001)Pugachev, Li, Boyer, Hancock, Le, Donaldson, and
      Xing]{Pugachev2001}
    Andrei Pugachev, Jonathan~G. Li, Arthur~L. Boyer, Steven~L. Hancock, Quynh~Thu
      Le, Sarah~S. Donaldson, and Lei Xing.
    \newblock {Role of beam orientation optimization in intensity-modulated
      radiation therapy}.
    \newblock \emph{International Journal of Radiation Oncology Biology Physics},
      50\penalty0 (2):\penalty0 551--560, 6 2001.
    \newblock ISSN 03603016.
    \newblock \doi{10.1016/S0360-3016(01)01502-4}.
    
    \bibitem[Rocha et~al.(2013)Rocha, Dias, Ferreira, and
      Lopes]{Rocha2013BeamMethod}
    Humberto Rocha, Joana~M. Dias, BrÃ­gida~C. Ferreira, and Maria~C. Lopes.
    \newblock {Beam angle optimization for intensity-modulated radiation therapy
      using a guided pattern search method}.
    \newblock \emph{Physics in Medicine and Biology}, 58\penalty0 (9):\penalty0
      2939--2953, 5 2013.
    \newblock ISSN 00319155.
    \newblock \doi{10.1088/0031-9155/58/9/2939}.
    
    \bibitem[Rocha et~al.(2018)Rocha, Dias, Ventura, Ferreira, and
      do~Carmo~Lopes]{Rocha2018}
    Humberto Rocha, Joana Dias, Tiago Ventura, BrÃ­gida Ferreira, and Maria
      do~Carmo~Lopes.
    \newblock {Comparison of combinatorial and continuous frameworks for the beam
      angle optimization problem in IMRT}.
    \newblock In \emph{Lecture Notes in Computer Science (including subseries
      Lecture Notes in Artificial Intelligence and Lecture Notes in
      Bioinformatics)}, volume 10961 LNCS, pages 593--606. Springer Verlag, 2018.
    \newblock ISBN 9783319951645.
    \newblock \doi{10.1007/978-3-319-95165-2{\_}42}.
    
    \bibitem[Rocha et~al.(2019)Rocha, Dias, Ventura, Ferreira, and Lopes]{Rocha}
    Humberto Rocha, Joana~Matos Dias, Tiago Ventura, BrÃ­gida da~Costa Ferreira,
      and Maria do~Carmo Lopes.
    \newblock {Beam angle optimization in IMRT: are we really optimizing what
      matters?}
    \newblock \emph{International Transactions in Operational Research},
      26\penalty0 (3):\penalty0 908--928, 5 2019.
    \newblock \doi{10.1111/itor.12587}.
    
    \bibitem[Romeijn et~al.(2005{\natexlab{a}})Romeijn, Ahuja, Dempsey, and
      Kumar]{Romeijn2005AModulation}
    H.~Edwin Romeijn, Ravindra~K. Ahuja, James~F. Dempsey, and Arvind Kumar.
    \newblock {A Column Generation Approach to Radiation Therapy Treatment Planning
      Using Aperture Modulation}.
    \newblock \emph{SIAM Journal on Optimization}, 15\penalty0 (3):\penalty0
      838--862, 1 2005{\natexlab{a}}.
    \newblock ISSN 1052-6234.
    \newblock \doi{10.1137/040606612}.
    \newblock URL \url{http://epubs.siam.org/doi/10.1137/040606612}.
    
    \bibitem[Romeijn et~al.(2005{\natexlab{b}})Romeijn, Kumar, Ahuja, and
      Dempsey]{Romeijn2005A}
    H~Edwin Romeijn, Arvind Kumar, Ravindra~K Ahuja, and James~F Dempsey.
    \newblock {A Column Generation Approach to Radiation Therapy Treatment Planning
      Using Aperture Modulation Railroad Scheduling View project Weapon Target
      Assignment View project A COLUMN GENERATION APPROACH TO RADIATION THERAPY
      TREATMENT PLANNING USING APERTURE MODULATION *}.
    \newblock \emph{Article in SIAM Journal on Optimization}, 2005{\natexlab{b}}.
    \newblock \doi{10.1137/040606612}.
    \newblock URL \url{http://www.siam.org/journals/siopt/15-3/60661.html}.
    
    \bibitem[Rowbottom et~al.(1999)Rowbottom, Webb, and Oldham]{Yan1999}
    Carl~Graham Rowbottom, Steve Webb, and Mark Oldham.
    \newblock {Beam-orientation customization using an artificial neural network}.
    \newblock \emph{Phys. Med. Biol}, 44:\penalty0 2251--2262, 1999.
    
    \bibitem[Rwigema et~al.(2015)Rwigema, Nguyen, Heron, Chen, Lee, Wang, Vargo,
      Low, Huq, Tenn, Steinberg, Kupelian, and Sheng]{Rwigema20154Toxicity}
    Jean Claude~M. Rwigema, Dan Nguyen, Dwight~E. Heron, Allen~M. Chen, Percy Lee,
      Pin~Chieh Wang, John~A. Vargo, Daniel~A. Low, M.~Saiful Huq, Stephen Tenn,
      Michael~L. Steinberg, Patrick Kupelian, and Ke~Sheng.
    \newblock {4{$\pi$} noncoplanar stereotactic body radiation therapy for
      head-and-neck cancer: Potential to improve tumor control and late toxicity}.
    \newblock \emph{International Journal of Radiation Oncology Biology Physics},
      91\penalty0 (2):\penalty0 401--409, 2015.
    \newblock ISSN 1879355X.
    \newblock \doi{10.1016/j.ijrobp.2014.09.043}.
    
    \bibitem[Sadeghnejad~Barkousaraie et~al.(2019)Sadeghnejad~Barkousaraie,
      Ogunmolu, Jiang, and Nguyen]{SadeghnejadBarkousaraie2019ATherapy}
    Azar Sadeghnejad~Barkousaraie, Olalekan Ogunmolu, Steve Jiang, and Dan Nguyen.
    \newblock {A Fast Deep Learning Approach for Beam Orientation Optimization for
      Prostate Cancer Treated with Intensity Modulated Radiation Therapy}.
    \newblock \emph{Medical Physics}, 12 2019.
    \newblock ISSN 0094-2405.
    \newblock \doi{10.1002/mp.13986}.
    
    \bibitem[Schreibmann and Xing(2005)]{Schreibmann2005}
    Eduard Schreibmann and Lei Xing.
    \newblock {Dose-volume based ranking of incident beam direction and its utility
      in facilitating IMRT beam placement}.
    \newblock \emph{International Journal of Radiation Oncology Biology Physics},
      63\penalty0 (2):\penalty0 584--593, 10 2005.
    \newblock ISSN 03603016.
    \newblock \doi{10.1016/j.ijrobp.2005.06.008}.
    
    \bibitem[Shirato et~al.(2018)Shirato, Le, Kobashi, Prayongrat, Takao, Shimizu,
      Giaccia, Xing, and Umegaki]{Shirato2018SelectionTreatment}
    Hiroki Shirato, Quynh~Thu Le, Keiji Kobashi, Anussara Prayongrat, Seishin
      Takao, Shinichi Shimizu, Amato Giaccia, Lei Xing, and Kikuo Umegaki.
    \newblock {Selection of external beam radiotherapy approaches for precise and
      accurate cancer treatment}.
    \newblock \emph{Journal of Radiation Research}, 59:\penalty0 i2--i10, 3 2018.
    \newblock ISSN 13499157.
    \newblock \doi{10.1093/jrr/rrx092}.
    
    \bibitem[Van't~Riet et~al.(1997)Van't~Riet, Mak, Moerland, Elders, and Van
      Der~Zee]{van1997conformation}
    Arie Van't~Riet, Ad~CA Mak, Marinus~A Moerland, Leo~H Elders, and Wiebe Van
      Der~Zee.
    \newblock A conformation number to quantify the degree of conformality in
      brachytherapy and external beam irradiation: application to the prostate.
    \newblock \emph{International Journal of Radiation Oncology* Biology* Physics},
      37\penalty0 (3):\penalty0 731--736, 1997.
    
    \bibitem[Ventura et~al.(2019)Ventura, Rocha, da~Costa~Ferreira, Dias, and
      Lopes]{Ventura2019ComparisonIMRT}
    Tiago Ventura, Humberto Rocha, Brigida da~Costa~Ferreira, Joana Dias, and Maria
      do~Carmo Lopes.
    \newblock {Comparison of two beam angular optimization algorithms guided by
      automated multicriterial IMRT}.
    \newblock \emph{Physica Medica}, 64:\penalty0 210--221, 8 2019.
    \newblock ISSN 1120-1797.
    \newblock \doi{10.1016/J.EJMP.2019.07.012}.
    \newblock URL
      \url{https://www.sciencedirect.com/science/article/pii/S1120179719301656}.
    
    \bibitem[Yarmand and Craft(2018)]{Yarmand2018EffectiveTherapy}
    Hamed Yarmand and David Craft.
    \newblock {Effective heuristics for beam angle optimization in radiation
      therapy}.
    \newblock \emph{Simulation}, 94\penalty0 (11):\penalty0 1041--1049, 11 2018.
    \newblock ISSN 17413133.
    \newblock \doi{10.1177/0037549718761108}.
    
    \bibitem[Yu et~al.(2018)Yu, Landers, Woods, Nguyen, Cao, Du, Chin, Sheng, and
      Kaprealian]{Yu2018}
    Victoria~Y. Yu, Angelia Landers, Kaley Woods, Dan Nguyen, Minsong Cao, Dongsu
      Du, Robert~K. Chin, Ke~Sheng, and Tania~B. Kaprealian.
    \newblock {A Prospective 4{$\pi$} Radiation Therapy Clinical Study in Recurrent
      High-Grade Glioma Patients}.
    \newblock \emph{International Journal of Radiation Oncology Biology Physics},
      101\penalty0 (1):\penalty0 144--151, 5 2018.
    \newblock ISSN 1879355X.
    \newblock \doi{10.1016/j.ijrobp.2018.01.048}.
    
    \bibitem[Yuan et~al.(2015)Yuan, Wu, Yin, Li, Sheng, Kelsey, and Ge]{Yuan2015a}
    Lulin Yuan, Q.~Jackie Wu, Fangfang Yin, Ying Li, Yang Sheng, Christopher~R.
      Kelsey, and Yaorong Ge.
    \newblock {Standardized beam bouquets for lung IMRT planning}.
    \newblock \emph{Physics in Medicine and Biology}, 60\penalty0 (5):\penalty0
      1831--1843, 2 2015.
    \newblock ISSN 13616560.
    \newblock \doi{10.1088/0031-9155/60/5/1831}.
    
    \bibitem[Yuan et~al.(2018)Yuan, Zhu, Ge, Jiang, Sheng, Yin, and Wu]{Yuan2018}
    Lulin Yuan, Wei Zhu, Yaorong Ge, Yuliang Jiang, Yang Sheng, Fang~Fang Yin, and
      Q.~Jackie Wu.
    \newblock {Lung IMRT planning with automatic determination of beam angle
      configurations}.
    \newblock \emph{Physics in Medicine and Biology}, 63\penalty0 (13), 7 2018.
    \newblock ISSN 13616560.
    \newblock \doi{10.1088/1361-6560/aac8b4}.

\end{thebibliography}
\bibliographystyle{plainnat}

\end{document}